\documentclass[useAMS,usenatbib,onecolumn]{mnras}
\pdfoutput=1

\usepackage{amsmath,amssymb}
\usepackage{graphicx}
\usepackage{color}
\usepackage{url}

\newcommand{\spose}[1]{\hbox to 0pt{#1\hss}}
\newcommand{\lta}{\mathrel{\spose{\lower 3pt\hbox{$\mathchar"218$}}
     \raise 2.0pt\hbox{$\mathchar"13C$}}}
\newcommand{\gta}{\mathrel{\spose{\lower 3pt\hbox{$\mathchar"218$}}
     \raise 2.0pt\hbox{$\mathchar"13E$}}}

\newcommand{\bI}{\mathbf{I}}
\newcommand{\bJ}{\mathbf{J}}
\newcommand{\bw}{\mathbf{w}}
\newcommand{\bl}{\mathbf{l}}
\newcommand{\bs}{\mathbf{s}}
\newcommand{\bx}{\mathbf{x}}
\newcommand{\bp}{\mathbf{p}}
\newcommand{\bO}{\mathbf{\Omega}}

\newcommand{\cC}{{\cal C}}

\title[Computational KAM I]{Practical application of KAM theory to
  galactic dynamics: I. Motivation and methodology}

\author[Weinberg]{Martin D. Weinberg\thanks{E-mail:
    weinberg@astro.umass.edu} \\
  Department of Astronomy, University of Massachusetts, Amherst,
}

\begin{document}

\label{firstpage}

\date{\today}
\pagerange{\pageref{firstpage}--\pageref{lastpage}} \pubyear{2015}

\maketitle

\begin{abstract}
  Our understanding of the mechanisms governing the structure and
  secular evolution galaxies assume nearly integrable Hamiltonians
  with regular orbits; our perturbation theories are founded on the
  averaging theorem for isolated resonances.  On the other hand, it is
  well-known that dynamical systems with many degrees of freedom are
  irregular in all but special cases.  For example, the solar system
  is chaotic and no doubt more so when dynamically young.  Similarly,
  although driven towards symmetry by dissipation and mixing galaxies
  are chaotic and, galaxies are dynamically young, are strongly
  affected by perturbations, and are probably strongly chaotic.  The
  best developed framework for studying the breakdown of regularity
  and the onset is the Kolmogorov-Arnold-Moser (KAM) theory.  Here, we
  use a numerical version of the KAM procedure to construct regular
  orbits (tori) and locate irregular orbits (broken tori).  Irregular
  orbits are most often classified in astronomical dynamics by their
  exponential divergence using Lyapunov exponents. Although their
  computation is numerically challenging, the procedure is
  straightforward and they are often used to estimate the measure of
  regularity.  The numerical KAM approach has several advantages over
  Lyapunov exponent computation: 1) it provides the morphology of
  perturbed orbits; 2) its constructive nature allows the tori to be
  used as basis for studying secular evolution; 3) for broken tori,
  clues to the cause of the irregularity may be found by studying the
  largest, diverging Fourier terms; and 4) it is more likely to detect
  weak chaos and orbits close to bifurcation.  Conversely, it is not a
  general technique and works most cleanly for small perturbations.
  We develop a perturbation theory that includes chaos by retaining an
  arbitrary number of interacting terms rather than eliminating all
  but one using the averaging theorem.  The companion papers show that
  models with significant stochasticity seem to be the rule, not the
  exception.  Moreover, the subdimensionally of phase-space features
  spawned by chaotic dynamics demands fine-scale temporal resolution,
  and we discuss the implication for n-body simulations.
\end{abstract}

\begin{keywords}
  methods: numerical --- dark matter --- galaxies: formation,
  evolution, kinematics and dynamics --- chaos
\end{keywords}

\section{Introduction}

\subsection{Overview}

Much of our insight into the dynamics of galaxy structure and
evolution relies on the properties of regular orbits.  We commonly
assume that regular orbits dominate owing to the importance of
baryonic dissipation during galaxy formation: dissipation naturally
selects non-intersecting orbits and is the principal enforcer of
axisymmetry.  Following the mostly successful $\Lambda$CDM paradigm,
we assume that galaxies are a combination of dissipative (disks) and
non-dissipative (dark-matter and stars) components with different
geometries.  There is no natural reason to assume that sufficient
conditions for regularity (i.e. the simultaneous separability of both
the Hamilton-Jacobi equation and the Poisson equation defined by the
St\"ackel conditions, e.g. \citealt{Vinti.etal:1998}) coincides with this
combination of dissipative and dissipationless dynamical processes.
Therefore, these dynamical systems are not regular everywhere almost
certainly.

On the other hand, owing to ease of computation, insight provided by
regular systems underpins most of our physical scenarios for galactic
stability, structure and evolution.  For example, theories of spiral
arms, swing amplification, among others assume regularity.  The author
has made predictions for the secular evolution of galaxies in combined
disk and spheroid systems using classical Hamiltonian perturbation
theory \citep[e.g.][]{Tremaine.Weinberg:84, Weinberg:98b,
  Weinberg:01c, Vesperini.Weinberg:01, Weinberg.Katz:02}.  This
previous work assumes that only one resonance influences the dynamics
of any one otherwise regular orbit at one time using the averaging
theorem \citep[e.g.][]{Arnold:89}.  The current paper assesses the
failure of this assumption and provides a tool for identifying
important dynamics missed by the averaging approximation.  Moreover,
environmental perturbations and secular and possibly dynamical
instabilities yield bars and spiral arms.  Some of these features are
strong perturbations and are likely to generate irregular orbits.
Specifically, is it possible that our ignorance of irregular behaviour
biases our interpretive insight?

In this paper, we present a general method based on one of the
mathematical tools used in some proofs of the Kolmogorov-Arnold-Moser
(KAM) theorem for studying the regular orbits that are preserved under
perturbations. Recall that regular orbits in a bound system are fully
characterised by a full set of constants of the motion, often chosen
to be action-angle variables for theoretical convenience.  These
variables are constructed by finding the canonical transformation from
a set of Cartesian variables (or some other conical coordinate system)
to a new set which leaves the Hamiltonian independent of angles. The
Hamilton-Jacobi (HJ) equation defines this canonical transformation
implicitly.  For regular systems, the action-angle variables define
tori in phase space that fully describe the quasi-periodic motion.
The approach pursued here is the construction of numerical solution of
the HJ equation for a perturbed regular system.  If the solution
exists, the new transformation yields a new set of action-angle
variables.  If the solution does not exist, the torus is presumed
destroyed; that is, one or more of the actions is no longer a constant
of the motion and the trajectory may appear chaotic.

This approach is complementary to the most popular approach for
studying chaos: Lyapunov exponent analysis.  While a study of Lyapunov
exponents provides a local phase-space measure for orbits whose
trajectories diverge exponentially under the influence of a
perturbation, exponent values close to zero provide little useful
information for the surviving regular orbits.  Conversely, the KAM
procedure characterises the orbits distorted by a perturber while
identifying but not characterising the orbit of the broken tori.
Finally, if irregular dynamics is important for galaxies, the dynamics
are likely to appear in stochastic layers around primary resonances
and these need to be accounted for by criteria used by n-body
simulators to determine phase-space and temporal resolution (see
\citealt{Springel:2005} for a widely-used example).  This paper
develops the basic formalism and presents the results for the circular
restricted three-body problem as an example.  In the companion papers,
we explore the question of how prevalent is irregularity in systems of
interest and how might this irregularity affect galaxy structure and
evolution?

Since these issues in modern non-linear dynamics will be unfamiliar to
many, we begin with a quick review of the theoretical motivation in
section \ref{sec:HJ} followed by a discussion of standard practice in
galaxy dynamics section \ref{sec:galdynintro}.

\subsection{Underlying dynamical theory}
\label{sec:HJ}

KAM theory is at the heart of modern Hamiltonian dynamics.  The
fundamental problem addressed by KAM theory first arose in the
three-body problem beginning with Poincar\'e: the introduction of a
third body into the integrable gravitational two-body system renders
the dynamics non-integrable, independent of the mass of the bodies.
There exists a large literature on perturbative tools for constructing
approximate integrable systems for solar-system dynamics
\citep[see][for a pre-KAM summary]{Brouwer.Clemence:61}.  These tools
are based on expansions the ratio of the planet mass to the solar
mass, a small parameter.  Then, they reduce the degrees of freedom of
the problem by identifying uninteresting degrees of freedom and
integrating or \emph{averaging} the resulting equations of motion over
these uninteresting variables.  However, the convergence of these
series is questionable owing to the appearance of small denominators
in the solution to the averaged equations.  Poincar\'e called this
fundamental problem of mechanics (see \citealt{Barrow:1997} for an
historical review).

The essence of the vanishing-denominator conundrum will be familiar to
most physics students in the context of a forced harmonic
oscillator. If the external force matches the oscillator's natural
frequency (i.e. is resonant) the linear solution is unbounded.  More
generally, the response of the non-linear oscillator is bounded but
will tend to generate oscillations in multiples of the driving
frequency resulting many resonances.  If we add a second non-linear
oscillator, the accidental near coincidence of two harmonics may lead
to erratic behaviour.  In a similar way, let us consider a solar
system with two planets; one planet exerts a periodic force on the
motion of a second.  Expanding the perturbation in a Fourier series,
we may recast the problem as that of coupled pendula forced by all
combinations of all possible harmonics of the natural frequencies of
the two planets.  The commensurability of the orbital periods for one
or more terms in the Fourier series of sufficient amplitude can lead
to small denominators resulting from resonant and divergence of the
expansion.

\citet{Kolmogorov:1954} suggested an two-part approach to obtaining
convergence of these series.  First, proceeding in the standard way,
one may linearize the problem about an approximate solution and obtain
an the approximate solution complete with small denominators.
Secondly, one may improve the approximate solution inductively using a
quadratically convergent iterative method.  Success of this approach
relies on the choice of initial conditions such that the denominators
do not vanish for low-order harmonic terms \emph{and} that the
amplitude of the harmonics terms approach zero exponentially with
increasing harmonic order.  These ideas were then made rigorous by
\citet[i.e.][]{Arnold:1963} and \citet{Moser:1962,Moser:1966} and led
to what we now know as the KAM theory.  In short, systems that can be
shown to satisfy the KAM theorem have to be sufficient smooth and
continuous and not have commensurate frequencies to some maximal order
for some perturbation amplitude $\epsilon$ smaller than some bound,
$\epsilon<\epsilon^\ast$.  

KAM theory does not solve the problem of non-integrability but helps
explain its genesis.  Specifically, tori may be destroyed if the
frequencies approach low-order commensurabilities or the perturbation
strength $\epsilon$ grows.  Although KAM theory does not estimate the
chaotic measure, direct numerical solution of the perturbed equations
of motion or evolution using Hamiltonian maps suggests that modest
perturbation strengths induce a non-zero measure of chaotic orbits.
The resulting phase space consists of a combination of regular and
irregular regions often called a \emph{mixed phase space}.  When most
tori are broken, even those far from the original location of the
vanishing denominators, the actions of the chaotic trajectories drift
in the irregular regions and the remaining tori are localised to
resonances, so-called \emph{regular islands}.  A detailed description
of this theory is beyond the scope of this paper.  However, while
reading various introductions to the current version of the KAM theory
\citep[e.g.][]{DeLaLlave:2003} while on sabbatical at the Institute
for Advanced Study several years ago, I was struck by the algorithmic
nature of Kolmogorov's underlying concept.  The basic tools are part
of the every numericist's standard arsenal: Newton's method and
numerical Fourier transforms.  Can one use these tools
\emph{numerically} to understand practical astronomical problems?  The
answer appears to be: YES!  The outstanding question for us here is:
does this generic dynamical picture of mixed phase space have
implications for galactic dynamics?

\subsection{Application to galactic dynamics}
\label{sec:galdynintro}

The same general ingredients necessary to understand modern celestial
dynamics applies to our understanding of galactic dynamics.  As in the
perturbation theories described in
\citet{Tremaine.Weinberg:84,Weinberg:94b}, we may exploit the
quasi-periodic nature of the unperturbed representation of the phase
space to convert the dynamical partial-differential equations to an
series of algebraic equations.  This series corresponds to an ordering
in increasing spatial frequencies.  Truncating this series on physical
grounds yields an algebraically tractable set of linear equations.
The main difference arises in the broader frequency spectrum and
therefore slower convergence the perturbed Kepler problem.  In
addition, the approach here abandons the averaging theorem that
artificially isolated resonances.

Using a related but orthogonal approach, James Binney and
collaborators have made enormous progress in their program to derive
approximate angle-action coordinates from the standard Cartesian
phase-space coordinates (and vice versa) that they call \emph{torus
  mapping}.  In a fiducial series of papers,
\citet{McGill.Binney:1990, Binney.Kumar:1993, Kaasalainen.Binney:1994}
show that generating functions formulated as action-angle series
together with a careful choice of a nearly analytic Hamiltonian that
approximates the system of interest and an appropriately tuned cost
function may be used to solve for the coefficients of the canonical
generating function with sufficient accuracy to obtain tori that are
excellent approximations to systems of astronomical interest.  More
recently, \citet{Sanders.Binney:2014} demonstrated that this approach
works well in three-dimensional as well as two-dimensional problems.

Although both approaches begin with an action-angle expansion of the
canonical generating function, we are attempting to discover, quantify
and understand the role of irregularity galaxies rather than trying to
obtain regular approximations to galaxies.  Torus mapping uses a cost
function to match the coefficients of the generating function series
in the least-squares sense to a target Hamiltonian.  In the approach
considered here, we attempt to implicitly solve for the values of
these coefficients that leaves the target Hamiltonian invariant.
Formally, this is equivalent to solving the Hamilton-Jacobi equation
that defines the actions and angles, as described earlier.

As a first galactic application, \citet[][hereafter Paper
2]{Weinberg:15b} applies the method developed in this paper to
investigate the role of chaos in barred galaxies.  Most certainly, we
are not the first to consider the importance of chaos in barred
galaxies; e.g. for a selection of relatively recent papers see
\citet[e.g.][]{Patsis:12, Manos.etal:13, Contopoulos.Harsoula:13,
  Manos.Machado:14, Ernst.Peters:14} and references therein.  Rather,
this work was intended to test the approach on a well investigated
astronomical scenario.  Using the new numerical KAM approach, we can
easily identify, in phase space, the tori that survive under the
self-consistent perturbation of the bar potential on the overall
gravitational potential.

Secondly, although collisionless and collisional dissipation lead to
systems that are approximately regular, the combination of the two are
\emph{not} regular.  Specifically, a dark halo yields a weakly
triaxial system, and therefore approximately spherical system.  Stable
spherical systems are regular.  The baryonic component quickly settles
into an approximately axisymmetric or elliptical disk, another nearly
regular system.  However the join of the two systems, as we envision
most galaxies, are unlikely to be regular.  Moreover, a general
three-dimensional stellar disk will not have regular
orbits. Therefore, there will be irregular, chaotic regions in phase
space in addition to regular regions. This simple observation raises
several questions.  Are there generic features of a exponential
stellar disk induced by breaking of invariant tori?  Do the chaotic
regions enable significant structural evolution in a galaxy lifetime?
Therefore, as a second application of the numerical KAM method, we
consider this problem in \citet[][hereafter Paper 3]{Weinberg:15c}.

\subsection{Plan for this paper}

To investigate these questions, we will develop a computational method
that seeks numerical solutions to the Hamilton-Jacobi (HJ) equation
for a irregular system by successive perturbations to an initially
regular system.  We will review the mathematical background and
develop the method in section \ref{sec:KAM}.  For the galactic
applications described above, we will begin with the spherical system
whose radial profile is obtained from the monopole or other
appropriate spherical approximation for gravitational potential of the
full non-spherical mass distribution.  The perturbation, then, is the
difference between full mass distribution and the monopole
distribution.  For example, to investigate irregularity of the
combined dark halo--galactic disk system, we consider the unperturbed
system to be the monopole part of the multipole expansion for the
entire system.  The resulting massless perturbation squeezes the
isopotential surfaces towards the disk plane.  Then, one may
investigate the evolving regular orbits (tori) by successively
increasing the perturbation until the target mass distribution is
reached.  section \ref{sec:HJ} describes the mathematical approach
followed by a description of the numerical algorithm whose details can
be found in Appendix \ref{sec:numerical}.  We conclude with a summary
and discussion in section \ref{sec:summary}.

\section{Using the KAM ideas numerically}
\label{sec:KAM}

\subsection{Introduction}
\label{sec:KAMintro}

The action-angle variables for a time-independent equilibrium system
are the result of seeking a canonical transformation that leaves the
Hamiltonian a function of momentum variables alone.  If the ansatz of
a generating function separable in the new momentum variables also
separates the Hamiltonian for bound values of the energy, the
resulting the resulting equations of motion are angle-like variables
that increase linearly with time describing the quasi-periodic motion.
The new momenta are constant in time because the Hamiltonian is
coordinate independent.  With an appropriate scaling, the new
coordinates may be chosen to advance by $2\pi$ for each cycle.  Let
these new momenta and coordinates (action-angle variables) be denoted
by $(\bI, \bw)$.  Geometrically, these new coordinates define a tori
in phase space.  If the system is everywhere regular, the phase space
is foliated by tori.

The existence of action-angle variables implies a system of regular
orbits.  Although this construction is sufficient but not necessary
for regularity, it is the usual starting point for most discussions.
More generally, orbits are often defined to be irregular if distance
between two phase-space states with small initial distance increases
exponentially.  This exponential divergence implies that any notion of
predictability or \emph{memory} of the initial state is soon lost.
Clearly, trajectories defined by action-angle variables are confined
to tori and will not be exponentially divergent.

Now, assume a system of regular orbits specified by action-angle
variables $(\bI, \bw)$ with the Hamilton $H_0(\bI)$. For the problems
considered in this paper, we will begin with a spherical unperturbed
mass distribution.  In the case, the values for $(\bI$, $\bw)$ are
always available by quadrature given $H_o(\bI)$.  Thus, the regular
system defined by a spherical potential provides a good starting point
for adding perturbations that induce irregularity.  The same features
hold true for any potential in St\"ackel form \citep{Stackel:1891} but
the general coordinate systems are more cumbersome
\citep[e.g.][]{Binney.Tremaine:2008}.  As we will see in the
development below, a chain of successive perturbations and canonical
transformations describing invariant tori will be tied, through
interpolation, to this fiducial foliation of phase space by regular
orbits.  We apply a non-spherical perturbation, $H_1(\bI, \bw) =
V_1(r,\theta,\phi)$ where $r, \theta, \phi$ are the radius,
colatitude, and azimuthal angle, respectively, so that the total
Hamiltonian is
\begin{equation}
  H(\bI,\bw) = H_0(\bI) + H_1(\bI, \bw).
  \label{eq:pertham}
\end{equation}
Given some non-spherical or non-axisymmetric perturbation, we often
rewrite equation (\ref{eq:pertham}) to \emph{define} the perturbation:
$H_1(\bI, \bw) \equiv H(\bI,\bw) - H_0(\bI)$.  For a spherical
distribution $\bI$ and $\bw$ are three-vectors \citep[see][for an
explicit representation]{Tremaine.Weinberg:84}.  We could use the
extended phase-space paradigm \citep[e.g.][Chap. 3]{Wiggins:2003} to
include time dependence without any substantive changes to this
development.  However, to simply notation, we assume that the
perturbation is time independent.

Following the standard procedure for solving the HJ equation, we seek
a new set of canonical variables, $(\bJ, \bs)$ say, such that
$H_{new}(\bJ) = H(\bI, \bw)$.  We assume a time-independent canonical
generator to transform from the original action-angle variables to a
new set of action-angle variables $(\bI, \bw)\xrightarrow{F_1(\bJ,
  \bw)}(\bJ, \bs)$ to first order in the perturbation:
\begin{equation}
  F_1(\bJ,\bw) \equiv \bJ\cdot\bw +
  W_1(\bJ,\bw).
  \label{eq:F1}
\end{equation}
This is a standard Type 2 generating function
\citep{Goldstein.etal:02}.  The first term in equation (\ref{eq:F1})
generates the identity transformation.  The new canonical variables
are then \citep[e.g.][\S48]{Arnold:89}
\begin{alignat}{2}
  \bI &= \frac{\partial F_1}{\partial\bw} = \bJ + 
  \frac{\partial W_1}{\partial\bw}, 
  &\qquad& \mbox{[old]} \label{eq:CI} \\
   \bs &= \frac{\partial F_1}{\partial\bJ} = \bw + 
   \frac{\partial W_1}{\partial\bJ} 
   && \mbox{[new]} \label{eq:Cw}
\end{alignat}
and the new Hamiltonian is
\begin{equation}
  H_{new}(\bJ) = 
  H\bigg(\bJ + \frac{\partial W_1}{\partial\bw}, \bw\bigg)
  =
  H_0\bigg(\bJ + \frac{\partial W_1}{\partial\bw}\bigg)
  + 
  H_1\bigg(\bJ + \frac{\partial W_1}{\partial\bw}, \bw\bigg).
  \label{eq:Hnew}
\end{equation}
Recall that
\begin{equation}
  H_{new}(\bJ) = E
  \label{eq:HJ}
\end{equation}
defines the time-independent HJ equation and is an implicit equation
for the term $W_1$ through equation (\ref{eq:CI}).  The key to finding
$H_{new}(\bJ)$ is the implicit solution for $W_1$ defined by equation
(\ref{eq:Hnew}).  Once we have the generating function $W_1$, the new
action variables follow immediately from equation (\ref{eq:CI}).

\subsection{The iterative canonical perturbation theory}
\label{sec:canonical}

A direct solution of the HJ equation (defined through equations
\ref{eq:Hnew} and \ref{eq:HJ}) using a numerical partial differential
equation scheme is not practical for problems of interest. Rather, we
define a solution based on a succession of first-order expansions.
The solution is then iterated to improve the approximation to $W_1$
until the desired convergence is obtained.

\subsubsection{Derivation of the algorithm}
\label{sec:derive}

The solution is constructed inductively as follows.  Assume that we
have an approximate solution $W_1^{[n]}$ at iteration $n$ whose
deviation from an exact solution is
\begin{equation}
  q^{[n]} \equiv H\bigg(\bJ+\frac{\partial W_1^{[n]}}{\partial\bw},
    \bw\bigg) - H_{new}^{[n]}(\bJ)
  \label{eq:q_n}
\end{equation}
where
\begin{equation}
  H_{new}^{[n]}(\bJ) \equiv \frac{1}{(2\pi)^3} \oint d\bw
  H\bigg(\bJ+\frac{\partial W_1^{[n]}}{\partial\bw}, \bw\bigg).
  \label{eq:Havg}
\end{equation}
The solution $W_1^{[n]}$ will be $\mathcal{O}{H_1}$, that is, the
order of the perturbing potential.  If $W_1^{[n]}$ is exactly $W_1$,
then equation (\ref{eq:q_n}) yields $q^{[n]}=0$ owing to equation
(\ref{eq:Hnew}).  Similarly, if $H_{new}^{[n]}(\bJ)$ is exactly
$H_{new}(\bJ)$ then the definition in equation (\ref{eq:Havg}) is an
identity.  However, if $q^{[n]}\not=0$, then $q^{[n]}$ is second-order
in the small perturbation $W_1^{[n]}$ by construction.  Now, let us
look for an improved solution, $W_1^{[n+1]} = W_1^{[n]} + \Delta
W_1^{[n]}$ where $\mathcal{O}[\Delta W_1^{[n]}] =
\mathcal{O}[q^{[n]}]$ from the definition in equation (\ref{eq:q_n}).
The improvement $\Delta W_1^{[n]}$ removes an additional piece of the
discrepancy between $H_{new}^{[n]}$ and $H_{new}$ which is already of
order $q^{[n]}$.  We will explicitly derive the improvement in
terms of the next order difference $q^{[n+1]}$ and demonstrate the
following ordering
\begin{equation}
  q^{[n+1]} =
  H\bigg(\bJ+\frac{\partial W_1^{[n+1]}}{\partial\bw}, \bw\bigg)
  -  H_{new}^{[n+1]}(\bJ) = \mathcal{O}[(q^{[n]})^2].
  \label{eq:qquad}
\end{equation}
This implies that the iteration converges quadratically. We assume the
existence of solution $W_1^{[n]}$ and endeavour to obtain the solution
$W_1^{[n+1]}$ such that
\begin{equation}
  H_{new}^{[n+1]}(\bJ) =  H\bigg(\bJ+\frac{\partial
      W_1^{[n+1]}}{\partial\bw}, \bw\bigg).
  \label{eq:iternp1}
\end{equation}
Although we do not know $H_{new}^{[n+1]}(\bJ)$ a priori, it is a
constant with respect to the new angle $\bs$ will not affect our
solution.  We begin with the Taylor series expansion of equation
(\ref{eq:iternp1}) in $\Delta W_1^{[n]}$:
\begin{eqnarray}
  H_{new}^{[n+1]}(\bJ) &=&   
  H\bigg(\bJ+\frac{\partial W_1^{[n+1]}}{\partial\bw}, \bw\bigg)
  = H\bigg(\bJ+\frac{\partial W_1^{[n]}}{\partial\bw} +
    \frac{\partial\Delta W_1^{[n]}}{\partial\bw}, \bw\bigg) \nonumber
  \\
  &=& 
  H\bigg(\bJ+\frac{\partial W_1^{[n]}}{\partial\bw}, \bw\bigg) +
  \frac{\partial H}{\partial\bI}\cdot\frac{\partial\Delta
    W_1^{[n]}}{\partial\bw} + \mathcal{O}[(q^{[n]})^2], \nonumber \\
  &=& 
  q^{[n]} + H_{new}^{[n]}({\bJ}) +
  \frac{\partial H}{\partial\bI}\cdot\frac{\partial\Delta
    W_1^{[n]}}{\partial\bw} + \mathcal{O}[(q^{[n]})^2]
  \label{eq:HJq}
\end{eqnarray}
having substituted equation (\ref{eq:q_n}) and $(\partial
H/\partial\bI)$ is evaluated at $\bI=\bJ+(\partial
W_1^{[n]}/\partial\bw)$.  The function $(\partial H/\partial\bI)$ may
be evaluated implicitly in terms of iteration quantities as follows:
\begin{eqnarray}
  \frac{\partial q^{[n]}}{\partial\bJ} = \frac{\partial H}{\partial\bJ} - 
  \frac{\partial H_{new}^{[n]}(\bJ)}{\partial\bJ}
  &=& \frac{\partial H}{\partial\bI}\cdot\bigg[\mathbf{1} + \frac{\partial^2
      W_1^{[n]}}{\partial\bJ\partial\bw}\bigg] - 
  \frac{\partial H_{new}^{[n]}(\bJ)}{\partial\bJ} \nonumber \\
  &=& \bigg[\mathbf{1} + \frac{\partial^2
      W_1^{[n]}}{\partial\bJ\partial\bw}\bigg]^T \cdot \frac{\partial
    H}{\partial\bI} - \frac{\partial H_{new}^{[n]}(\bJ)}{\partial\bJ}.
\end{eqnarray}
After rearranging, we find
\begin{equation}
  \frac{\partial H}{\partial\bI} = \bigg(
    \bigg[\mathbf{1} + \frac{\partial
        W_1^{[n]}}{\partial\bJ\partial\bw}\bigg]^T\bigg)^{-1}
  \bigg(\frac{\partial q^{[n]}}{\partial\bJ} + 
    \frac{\partial H_{new}^{[n]}}{\partial\bJ}\bigg).
  \label{eq:dHdI}
\end{equation}
Since
\begin{equation}
  \bigg[\bigg(\mathbf{1} + \frac{\partial^2
        W^{[n]}_1}{\partial\bJ\partial\bw}\bigg)^T\bigg]^{-1} 
  \bigg(\frac{\partial q^{[n]}}{\partial\bJ}\bigg) \cdot
  \frac{\partial\Delta W^{[n]}_1}{\partial\bw}
  = \mathcal{O}\bigg[(q^{[n]})^2\bigg],
\end{equation}
we may ignore the $\partial q^{[n]}/\partial\bJ$ term in equation
(\ref{eq:dHdI}) upon substituting into equation (\ref{eq:HJq}) to
$\mathcal{O}[(q^{[n]})^2]$.  Performing the substitution, the equality
in the last of equation (\ref{eq:HJq}) becomes
\begin{equation}
  q^{[n]} +
  \frac{\partial H_{new}^{[n]}}{\partial\bJ}\cdot
  \bigg[\bigg(\mathbf{1}+\frac{\partial^2
        W_1^{[n]}}{\partial\bJ\partial\bw}\bigg)^T\bigg]^{-1}
  \cdot\frac{\partial\Delta W_1^{[n]}}{\partial\bw} =
  H_{new}^{[n+1]}(\bJ) - H_{new}^{[n]}(\bJ)
  \label{eq:solHJ}
\end{equation}
up to terms with $\mathcal{O}[(q^{[n]})^2]$.  Because $q^{[n]}$,
$H_{new}^{[n]}$ and $W_1^{[n]}$ are known quantities at this point in
the iteration, equation (\ref{eq:solHJ}) may be solved explicitly for
$\Delta W_1^{[n]}$ to first-order accuracy yielding a solution to
$q^{[n+1]}=0$ that is accurate to $\mathcal{O}[(\Delta q^{[n]})^2]$.
In other words, we have shown that $\mathcal{O}[q^{[n+1]}] =
\mathcal{O}[(\Delta q^{[n]})^2]$, and, therefore, the iterative
correction to $q^{[n]}$ defined by equations (\ref{eq:q_n}) and
(\ref{eq:solHJ}) converges quadratically.

A compact solution of equation (\ref{eq:solHJ}) is obtained by Fourier
analysis in the new angles $\bs$ using the standard Type 2 canonical
transformation (equation \ref{eq:Cw}).  We assume that the
action-angle expansion of the generating function $W_1$ in the new
action-angle variables exists:
\begin{equation}
  W_1 = \sum_{\bl} \varpi_{\bl}(\bJ) e^{i\bl\cdot\bs}.
  \label{eq:w1aa}
\end{equation}
The Fourier transform of equation (\ref{eq:solHJ}) may be written
\begin{eqnarray}
  q_{\bl}&=&
  \frac{1}{(2\pi)^3}\oint d\bs q^{[n]}(\bJ,\bw(\bJ,\bs))
  e^{-i\bl\cdot\bs} \label{eq:ql1} \nonumber \\
  &=& \frac{1}{(2\pi)^3}\oint d\bs
  \bigg\{
    \frac{\partial H_{new}^{[n]}}{\partial\bJ}\cdot
    \bigg[\bigg(\mathbf{1}+\frac{\partial^2
          W_1^{[n]}}{\partial\bJ\partial\bw}\bigg)^T\bigg]^{-1}
    \cdot\frac{\partial\Delta W_1^{[n]}}{\partial\bw}
  \bigg. \nonumber \\
  && - \bigg(H_{new}^{[n+1]}(\bJ) - H_{new}^{[n]}(\bJ)\bigg)
    \Bigg\}
  e^{-i\bl\cdot\bs}.
  \label{eq:ql2}
\end{eqnarray}
The first term in the integrand of equation (\ref{eq:ql2}) may be
simplified using the canonical transformation (equation \ref{eq:Cw}):
\begin{eqnarray}
  \bigg[\bigg(\mathbf{1}+\frac{\partial^2
        W_1}{\partial\bJ\partial\bw}\bigg)^{T}\bigg]^{-1} \cdot
  \frac{\partial\Delta W_1}{\partial\bw} &=&
  \bigg[\mathbf{1} + \frac{\partial^2
      W_1}{\partial\bw\partial\bJ}\bigg]^{-1} \cdot
  \frac{\partial\Delta W_1}{\partial\bw} \nonumber \\ &=& 
  \bigg(\frac{\partial\bs}{\partial\bw}\bigg)^{-1}\cdot
  \frac{\partial\Delta W_1}{\partial\bw} = 
  \frac{\partial\Delta W_1}{\partial\bs}.
\end{eqnarray}
The second term in the integrand is independent of $\bs$ and therefore
vanishes unless $\bl=0$.  Finally, in terms of the Fourier
coefficients, the solution to equation (\ref{eq:solHJ}) is
\begin{equation}
  q_{\bl} + (H_{new}^{[n]} - H_{new}^{[n+1]})\delta_{\bl\, {\bf 0}} +
  i\bl\cdot\frac{\partial H_{new}^{[n]}}{\partial\bJ}\Delta\varpi_{\bl} = 0
\end{equation}
or, for $\bl\not=0$,
\begin{equation}
  \Delta\varpi_{\bl}(\bJ) = \frac{i}{\bl\cdot\frac{\partial
      H_{new}^{[n]}}{\partial\bJ}} q_{\bl}.
  \label{eq:ql3}
\end{equation}
Since an arbitrary additive constant in the generating function $W_1$
does not affect the solution for the action-angle variables, we are
free to ignore the constant coefficients with $\bl=0$.  Repeating the
procedure, we may continue to iterate until $q^{[n]}$ is as small as
desired.  This iteration method is essentially that used in proving
the KAM theorem \citep[][Chapter 5, \S2]{Arnold.etal:97}.  As in this
theorem, the iterative corrections are second-order in the
perturbation and are, therefore, quadratically convergent.  From
Hamilton's equations, the quantity $\mathbf{\Omega}=\partial
H_{new}^{[n]}/\partial\bJ$ is a vector of frequencies and the small
divisor $\bl\cdot\mathbf{\Omega}$ will occur near a commensurability
or resonance.  Even when the denominator is bounded away from zero,
the effect of the denominator is far reaching (see section
\ref{sec:KAMdiscuss} below for discussion).

This procedure has some favourable features.  The iteration produces
the series in the new actions and angles directly.  Also, assuming
that the Hamiltonian is available, the computation of the integrands
in equation (\ref{eq:ql1}) do not require gradients in $\bJ$; that is,
the iteration can proceed without computing the frequencies $\partial
H/\partial\bJ$.  The updated coefficients are provided recursively by
$\varpi^{[n+1]}_{\bl}(\bJ) \leftarrow \varpi_{\bl}^{[0]}(\bJ) +
\sum_{k=1}^{n}\Delta\varpi_{\bl}^{[k]}(\bJ)$ and $H^{[n+1]}_{new}
\leftarrow H_0 + \sum_{k=1}^{k} q_{\bl=\mathbf{0}}^{[k]}$ where the
initial value of $q^{[1]}$ is simply the first order perturbation (see
also equation \ref{eq:HJ1st}).  There are, however, several practical
problems with the formulation above.  The main one is that evaluation
of equation (\ref{eq:ql1}) requires $\bw=\bw(\bJ, \bs)$ which is
implicit.  This solution be computed straightforwardly using the
Newton-Raphson method.  An alternative approach uses an expansion in
the new actions $\bJ$ and old angles $\bw$ The mixed solution avoids
an implicit solution of equation (\ref{eq:Cw}) to obtain the old
angles in terms of the new.  A rederivation of the iterative solution
for the mixed Fourier variables that yields a numerically expedient
algorithm for the computation of the expansion coefficients using the
standard Fast Fourier Transform algorithm without implicit solutions
is presented in Appendix \ref{sec:numerical}.  Each choice has
advantages depending on the application.

Finally, the quadratic convergence of the iteration suggests a direct
Newton-Raphson evaluation of equation (\ref{eq:solHJ}).  While
possible, numerical stability seems to require a careful choice of a
small but dominating subset of indices $\bl$.  Since our goal is an
automated analysis, we used equation (\ref{eq:ql3}) or equation
(\ref{eq:HJN2}) for the examples in this and in companion papers.

\subsubsection{Discussion}
\label{sec:KAMdiscuss}

Mathematically speaking, a term in equation (\ref{eq:w1aa}) based on
the solution from equation (\ref{eq:ql3}) with a vanishing denominator
will result in long-period motion with large amplitude for one degree
of freedom. This signals the breakdown of the continuous, predictable
behaviour of linear perturbation theory.  The KAM theory arose as a
result of Poincar\'e's concern about the stability of the three-body
problem (i.e. the stability of the solar system) resulting from this
breakdown.  The two systems have one huge qualitative difference:
dynamical times in the solar system are measured in the billions while
dynamical times in a galaxy are measured in the hundreds, maybe
thousands at most.  Owing to our mathematical techniques, the
numerical characterisation of the perturbation theory failure often
requires extremely high-resolution data and time series very long
compared to the natural frequencies of our dynamical system. This is
partly the result of our celestial mechanics bias and computational
limitations.  Galaxies have \emph{many} orders of magnitude more stars
and dark-matter particles than solar system bodies of interest.
Therefore, even though the available time scales are much shorter, the
phase space coverage is so thorough that some an observable subset of
particles will manifest chaotic dynamics in the time available.  This
regime begs for study and partially motivates this paper.

Much of KAM theory is concerned with the phase-space domain that
admits convergent solutions away from the vanishing denominators in
equation (\ref{eq:ql3}).  The theory employs the exponential decay of
Fourier coefficients of analytic functions with increasing $|\bl|$
together with the general condition that rational numbers approximate
the ratios of Hamiltonian frequencies, $\partial H/\partial\bI$,
sufficiently badly that the series converges nearly everywhere for
most problems of physical interest under weak perturbations.  This
condition is often written in the following form:
\begin{equation}
  \bl\cdot\frac{\partial H_{new}^{[n]}}{\partial\bJ} \ge
  \beta (|\bl|)^{-\gamma}
  \label{eq:diophantine}
\end{equation}
for some $\beta>0$ and $\gamma\ge d$ where $d$ is the rank of $\bl$.
Armed with this formal result, we have confidence that our procedure
will work much for much of phase space, although we will not
explicitly verify the exponential decrease of coefficient amplitudes
here.  However, note that the time-scale mismatch remains; the eternal
nature of the torus may be too strong a criterion for our galactic
systems.  We may address this in an intuitive way by noting that the
time-scale for a particular term with a vanishing denominator will be
proportional to the maximum absolute value of the integers in $\bl$
(denoted $||\bl||$).  Therefore, a truncation at modest values of
$||\bl||$ combined with the exponential decay of high-order
coefficients suggests that the dynamics revealed by the algorithm from
section \ref{sec:derive} may retain its applicability.  We will see
examples of this truncation in section \ref{sec:3body}

In the standard application of the KAM theorem, one attempts to find a
torus subject to the constraint of equation (\ref{eq:diophantine}).
This may require some adjustments to the initial conditions in $\bI$.
Our goal here is somewhat different: we are interested in the fraction
of tori that break up under the perturbation as a function of phase
space volume and therefore, we do not adjust $\bI$. In other words, we
are interested in the converse to the KAM problem: where in phase
space do tori \emph{not} exist?  The natural implication of failing
convergence of the series $q_{\bl}(\bJ)$ for some $\bJ$ is a broken
torus. In practice, failure has two forms: 1) the quantity
$\bl\cdot\frac{\partial H_{new}^{[n]}}{\partial\bJ}=\epsilon\sim0$ so
that the amplitude $\Delta W_{\bl}$ is too large to continue with the
iteration; and 2) multiple terms of different $\bl$ influence each
near one of the separatrices, inducing Smale horseshoes (and chaos
thereby) into the dynamics around the homoclinic trajectory \citep[as
in][]{Holmes.Marsden:82}.  Although the KAM theorem does not offer
specific wisdom on this, numerical experiments described below and in
\citetalias{Weinberg:15b} and \citetalias{Weinberg:15c} show that
failure to converge is most frequently an indicator of chaos.

\subsection{Incremental perturbations}

The KAM procedure is perturbative.  Some applications, therefore, may
benefit from breaking a large perturbation into small increments to
better understand the development of irregularity.  For example, we
might choose to define a perturbation strength ${\bar\epsilon}_j =
j\epsilon/m$ with $j\in[1,m]$.  Then, at increment $j$, the
perturbation strength is always of order $\epsilon/m$ with
\begin{eqnarray}
  H_0(j) &\equiv& H_0 + \frac{j-1}{m} H_1, \label{eq:incrp} \\
  H_1(j) &\equiv& \frac{1}{m}H_1,
\end{eqnarray}
with
\begin{equation}
  H(j) = H_0(j) + H_1(j).
\end{equation}
Clearly, this could be generalised to a non-linear dependence on
$\epsilon$.  Then, the algorithm from section (\ref{sec:derive}) may
be applied successively to each level $j$.  Explicitly, let the
solution at each step $j$ be denoted by $W_1(\bJ,\bw; j)$, or using
equation (\ref{eq:F1}), $F_1(\bJ,\bw; j) = \bJ\cdot\bw + W_1(\bJ,\bw;
j)$.  The iteration developed in the previous section results in the
following chain of canonical transformations:
\begin{eqnarray}
[\bI, \bw] &\equiv& [\bI(0), \bw(0)]
\xrightarrow{F_1(\bJ,\bw;1)}
[\bI(1), \bw(1)]
\xrightarrow{F_1(\bJ,\bw;2)}
[\bI(2), \bw(2)]\cdots \nonumber \\
&& [\bI(m-2), \bw(m-2)]
\xrightarrow{F_1(\bJ,\bw;m-1)}
[\bI(m-1), \bw(m-1)]
\xrightarrow{F_1(\bJ,\bw;m)}
\nonumber \\ &&
[\bI(m), \bw(m)] \equiv [\bJ, \bs].
\label{eq:chain}
\end{eqnarray}
At any step in the chain, the converged perturbed orbit has
$\bI(j)=\mbox{constant}$ with uniformly advancing $\bw(j)$ at the rate
$\bO(j)\equiv\partial H(\bI(j);j)/\partial\bI(j)$.  

For a simple example, consider a chain with a single link, $m=1$.  We
wish to plot the perturbed trajectory for the torus with $\bJ$ as a
function of time.  By construction, the angle variable advance
uniformly at the rate $\mathbf{\Omega}$, so beginning at $\bs(0)$,
choose to advance to $\bs(h) = \bs(0) + \mathbf{\Omega}h$.  Using
equations (\ref{eq:CI}) and (\ref{eq:w1aa}), we compute $\bI$ at
$\bs(t=h)$.  Using equations (\ref{eq:Cw}) and (\ref{eq:w1aa}), we
compute for $\bw$ at $\bs(t=h)$.  Since the Cartesian values of the
phase-space point $(\bx, \bp)$ are known since the regular orbit is
determined by quadrature, we have our phase space point for the torus
$\bJ$ at $t=h$.  For $m>1$, the same steps are used recursively from
beginning at $j=m$ and moving down the chain by recursion.  Equation
(\ref{eq:Cw}) tells us that the $\bw(j-1)$ is implicitly defined by
$[\bI(j), \bw(j)]$.  Solving for $\bw(j-1)$, equation (\ref{eq:CI})
gives $\bI(j-1)$ explicitly in terms of $\bI(j)$. We then walk down
the chain transformations defined by equation (\ref{eq:chain}) for
each desired value of $[\bI(j), \bw(j)]$ until we reach $[\bI,
\bw]\equiv[\bI(0), \bw(0)]$.  The last step provides us with the
phase-space point in terms of the unperturbed tori.  The penalty of
using equations (\ref{eq:incrp})--(\ref{eq:chain}) for large $m$ is
the large memory needed to store all of the coefficients and
intermediate tables of actions and angles at each $j$. See
\citetalias{Weinberg:15c} for additional discussion of this topic and
its application.

\subsection{Numerical considerations}

\subsubsection{Numerical partial derivatives}
\label{sec:interp}

The computation of new perturbed action-angle values along with the
computation of the new frequencies requires action-space partial
derivatives of the expansion coefficients from equations
(\ref{eq:ql1}) and (\ref{eq:ql3}) using equations (\ref{eq:CI}),
(\ref{eq:Cw}) and (\ref{eq:Hnew}). These numerical computations rely
on either finite-difference or functional approximation techniques,
and as usual, the best choice is problem dependent. We will use the
notation defined in \citet{Tremaine.Weinberg:84} for action variables
in a spherically-symmetric gravitational potential: $I_1$ and $I_2$
denote the radial and azimuthal action in the orbital plane, and $I_3$
denotes the projection of $I_1$ on the axis perpendicular to the
equatorial plane (i.e. the $z$ axis).  Depending on the perturbation,
the perturbed actions $J_1, J_2, J_3$ will no longer have these clear
interpretations.  For a perturbation to a thin disk by a vertically
symmetric disturbance, we can restrict our consideration to the action
pair $(I_1, I_2)$ and consider only a two-dimensional distribution of
action points. In general, an initially rectangular grid of nodes
$(I_1, I_2)$ will not rectangular after the perturbation is applied.
Therefore, we can not rely on the standard algorithms that assume a
rectilinear grid for $(J_1, J_2)$ for interpolation.  For this paper,
we explored two techniques to handle the expected irregular
distribution of nodes $(J_1, J_2)$.  First, the modified Shepard
algorithm \citep{Shepard:68} uses a Voronoi tessellation to provide a
metric on the unstructured grid and inversely weights the contribution
of nearby points by distance within a specified radius.
Alternatively, one may use a k-d tree to identify points within the
specified radius using a nearest-neighbour search algorithm.  The
latter is slightly more efficient in greater than two dimensions.
Secondly, \citet{Riachy.etal:2011} find an optimal set of coefficients
in the least-squares sense to a multidimensional Jacobi polynomial
approximation in an irregular region.  The Shepard algorithm performed
more robustly in tests and we use this approach for this and the
companion papers.

The third action, $I_3$, is conserved for an axisymmetric but
otherwise three-dimensional perturbation (such the effect of a
flattened perturbation on spherical bulge and halo).  Therefore, we
may consider slices of constant $I_3$ individually and break the
problem into two pieces.  We first define planes of constant
$I_3=L_z$.  The maximum value of $I_3$ may be defined by choosing an
effective circular radius: $r_{circ}$ and setting $I_3 = r_{circ}
v_{circ}$.  Since $I_2\ge I_3$ and $I_3 = I_2\cos(\beta)$ where
$\beta$ describes the inclination of the orbital plane relative to the
$z=0$ plane, once the value of $I_3$ is fixed, the values of $I_2>I_3$
determine $\beta$.  In other words, the orbits defined on the
$I_1$--$I_2$ plane at fixed $I_3$ have larger inclination off the disk
plane as $I_2$ increases. This is a slightly odd way of representing
phase space but is convenient for numerical derivatives.  For example,
$\Omega_3=\partial H(\bI)/\partial I_3$ may be computed by
interpolating $H$ for the desired values of $(I_1, I_2)$ and using
finite-difference derivative between $I_3$ planes to compute $\partial
H(\bI)/\partial I_3$.  Because $I_2 \ge |I_3|$, the computation of
derivatives at the lower boundary in $I_2$ may require special
treatment such as extrapolation.

\subsubsection{Computing the action and angle transforms}
\label{sec:actang}

We truncate the action-angle series by choosing a maximum index in
each dimension, $l_{j,max}$.  These indices should be chosen
sufficiently large to obtain an accurate representation of the orbit
by the trigonometric polynomial.  For eccentric orbits, whose
expansions include terms with large $||\bl||$ owing to the high
instantaneous frequencies at pericentre, the required value of
$l_{j,max}$ may be much larger than the desired number of retained
commensurabilities.  However, overly large values of $l_{j,max}$ are
computationally costly, both in CPU time and memory usage.  For
computations here, we find that $l_{j,max}\le16$ is a good compromise.
We use the efficient FFTW package\footnote{\url{http://www.fftw.org/}}
for the computing the discrete Fourier transform using the
fast-Fourier transform algorithm.

One of the pair of transform equations (equations \ref{eq:CI} and
\ref{eq:Cw}) is implicit, depending on whether we use the angle
expansion in new variables (as in equation \ref{eq:w1aa}) or in old
angle variables (as described in Appendix \ref{sec:numerical},
equation \ref{eq:Wexp}).  As described at the close of section
\ref{sec:derive}, the mixed variable form that follows the Type 2
generating function explicitly is computationally more efficient.  In
this case, transforming from the new action to the old action given
the old angle is a simple application of equation (\ref{eq:CI}) given
$\varpi_{\bl}$ from equation (\ref{eq:HJN}) or equation
(\ref{eq:HJN2}):
\begin{equation}
  \bI = \bJ + 
  \frac{\partial W_1}{\partial\bw} = 
  \bJ + i \sum_{\bl} \bl\,\varpi_{\bl}(\bJ)
e^{i\bl\cdot\bw}.
\end{equation}
The angle transform is more complicated in two ways.  First, equation
(\ref{eq:Cw}) requires an action derivative of $\varpi_{\bl}(\bJ)$,
which we perform using the modified Shepard algorithm as described in
section \ref{sec:interp}.  Second, equation (\ref{eq:Cw}) is now an
implicit expression for $\bs$ in terms of $\bw$: $\bs=\bs(\bJ,\bw)$.
Fortunately, this is efficiently solved using the Newton-Raphson
method as follows.  Define:
\begin{equation}
  \mathbf{\cal Q}(\bw) = \bw + 
  \sum_{\bl} \frac{\partial\varpi_{\bl}}{\partial\bJ}
    e^{i\bl\cdot\bw}
    \label{eq:calQ}
\end{equation}
and
\begin{equation}
  \mathbf{\cal M}(\bw) \equiv \frac{\partial\mathbf{\cal
      Q}}{\partial\bw} = \mathbf{1} + 
  i \sum_{\bl} \bl\,\frac{\partial\varpi_{\bl}}{\partial\bJ} e^{i\bl\cdot\bw}
  \label{eq:calM}.
\end{equation}
Then, the Newton-Raphson iteration for $\bw$ becomes
\begin{equation}
  \bw_j = \bw_{j-1} +
  {\cal M}^{-1}\cdot\bigg(\bs - {\cal Q}(\bw_{j-1})\bigg).
\end{equation}
This converges to machine precision in under ten iterations in practice.

\subsubsection{Time dependence}
\label{sec:timedep}

The formalism of section \ref{sec:canonical} may easily be extended to
time-dependent perturbations.  Similar to the transformation of the HJ
equation to an algebraic equation using action-angle transforms, we
may treat the time variable using a Laplace transform.  The special
case of a simple harmonic time dependence, such as that resulting from
a constant pattern speed with frequency $m\Omega_p$, a Fourier
transform is sufficient.  The resulting transform has a single term
and the denominators in equation (\ref{eq:ql3}) become:
\begin{equation}
  \bl\cdot\frac{\partial H_{new}^{[n]}}{\partial\bJ} \longrightarrow
  \bl\cdot\frac{\partial H_{new}^{[n]}}{\partial\bJ}  - m\Omega_p.
\end{equation}
We sum over all driving frequencies $j$ if there are a finite number
of discrete constant drivers $m_j\Omega_{p\,j}$.  If the
time-dependence is aperiodic, the resulting transform results in a
continuous frequency-like Laplace transform variable $s$:
\begin{equation}
  \bl\cdot\frac{\partial H_{new}^{[n]}}{\partial\bJ} \longrightarrow
  \bl\cdot\frac{\partial H_{new}^{[n]}}{\partial\bJ}  + i s.
\end{equation}
The inverse Laplace transform required for the fully general
time-dependent Hamiltonian requires an additional grid of quadratures
but otherwise, the algorithm follows as in section \ref{sec:canonical}.
In the example below and in \citetalias{Weinberg:15b} and
\citetalias{Weinberg:15c}, we will assume at most a single driving
frequency.
  
\section{Software implementation}

\subsection{Tori computation}

\label{sec:toricomp}

The ideas from the previous sections have been implemented in an
object-oriented C++ code.  Readers with little interest or experience
in object-oriented programming (OOP) may skip this section without
consequence. In short, an \emph{object} or \emph{class} is a data
structure and a set of functions.  OOP imbues the relationship between
be classes with the same relationships and interrelationships that
their counterparts have in the physical or mathematical world.  Thus,
OOP provides a framework for the computational scientist to build and
extend their numerical machinery (see \citealt{Pitt-Francis:2012} for
an illustrative introduction).  Because OOP enforces the natural
relationships between objects, it also helps prevent mistakes.  The
specific OOP design features for the code that implements the
algorithm in section \ref{sec:derive} are as follows:
\begin{enumerate}
\item The Hamiltonian torus concept is embodied by a single class
  whose instances track the action vector, the coefficients of the
  $W_1$ expansion, and the derivatives of $W_1$ necessary to compute
  the angles at any desired time and at any level of the perturbation
  hierarchy.
\item Time-dependent perturbations will require multiple harmonic
  orders or frequency subspaces.  These may be used to track periodic
  or aperiodic temporal perturbations using Fourier or Laplace,
  respectively.  These subspace domains are fixed by the definition of
  the perturbation.  The torus class instances keep lists of values
  indexed by each subspace.
\item A container class keeps track of the full collection of
  action-vector \emph{nodes} and associated tori for each
  perturbation.  This container implements the interpolation and
  differentiation of the possibly irregularly sampled action points
  that enables the computation of the perturbed trajectories for each
  new torus.
\item This container class may have any number of derived classes that
  may initialise the individual nodes with action values based on
  specific symmetries or special geometries in phase space.  Since the
  calculation begins with regular orbits, a particular derived class
  implements the computation of the analytic mapping between Cartesian
  and angle-angle phase-space vectors for each geometry of interest.
  The astronomical applications here and in the companion papers
  assume axisymmetric or spherical unperturbed models.
\item The container class of nodes has been designed to handle a
  three-dimensional unstructured grids, two-dimensional grids, and
  three-dimensional grids restricted to two-dimensional slices with
  fixed $I_3$. During development, we tested the general
  three-dimensional grid using a computation that was restricted to
  two-dimensional slices.  As expected, the implementation of a full
  three-dimensional grid is the most expensive and noisiest, requiring
  an order of magnitude more tori to achieve the same accuracy than
  the equivalent restriction to two-dimensional slices.
\item The container class keeps track of and attempts to
  diagnose any problems in the interpolations as the computation
  proceeds.  Specifically, for strong perturbations, the induced
  oscillations for one $\varpi_\bl$ term on another may drive
  excursions beyond the initial action grid.  Most often, these happen
  near phase-space boundaries for diverging iterations typical of
  broken tori.
\item Finally, these computations often represent an investment in
  both human and computing time.  All intermediate results are stored
  to full precision for later reuse using the BOOST
  (\url{http://www.boost.org}) serialisation library.
\end{enumerate}

For each newly-defined computation, we begin by choosing an
appropriate value of $l_{j,max}$ for each dimension in the subspace.
This finite set of ``quantum numbers'' results in a cubical lattice of
$(l_1, l_2, l_3)$.  For two-dimensional problems, the lattice is a
square.  We have explored the dependence of the coefficients on the
selected value by increasing $l_{max}$ by factors of two until
convergence reached.  For problems explored to date $l_{max}\le16$ has
been sufficient.  Also, significantly larger values require special
accommodations memory requirements.

\subsection{Computation of Lyapunov exponents}
\label{sec:lyapcompute}

The Lyapunov exponents, $\lambda_j$ for $j=0,\ldots,n-1$ where $n$ is
phase-space dimensionality, measure the exponential divergence rate of
initially nearby phase-space points. The so-called \emph{standard
  algorithm} \citep{Wolf.etal:85} exploits the multiplicative ergodic
theorem \citep{Oseledets:68}.  The import of this theorem may be
motivated by the following example.  Consider the mapping defined by
the continuous solution of the linearized equations of motion,
$\mathbf{\dot u} = \mathbf{G}(\mathbf{x}, t) \cdot \mathbf{u}$ where
$G_{ij} = \partial F_i/\partial x_j$.  This gives us a tangent-space
mapping from $t=0$ to $t=h$: $\mathbf{u}(h) =
\mathbf{T}_h[\mathbf{x}(0)]\mathbf{u}(0)$.  If our tangent map
$\mathbf{T}_h$ were constant for all time, it would be a trivial
matter to compute the eigenfunctions and eigenvalues once, and compute
the subsequent evolution at any desired time.  We cannot do this
because $\mathbf{T}$ is time dependent, and therefore we would have a
new eigenvalue problem at every time $t$.  Oseledets' theorem tells us
that the generic case behaves as if $\mathbf{T}$ were constant and
that the Lyapunov exponents are related to the eigenvalues of the the
mapping matrix: $\lambda_j=\lim_{t\rightarrow\infty} t^{-1}
\log(|\mathbf{T}\cdot\mathbf{u}_j|/|\mathbf{u}_j|)$ for appropriately
chosen eigenvectors $\mathbf{u}_j$.

The major shortcomings of the algorithms based on this theorem are
that the accuracy of the estimates depend sensitively on the numerical
method.  Because of this, there is a very large literature describing
improved Lyapunov exponent algorithms.  However, for our purposes
here, accurate exponents are not required.  We will use only the
standard algorithm to discriminate between exponents that exceed some
positive threshold value and those that do not.  Also, following
common practice, we will use the continuous Gram-Schmidt normalisation
procedure \citep{Christiansen.Rugh:97} to estimate the various
exponential subspaces to prevent the computational singularity that
results from the direct application of the Oseledets theorem.  There
are many newer approaches with virtues for different applications
\citep[see][for a review]{Darriba.etal:2012}; the choice of the
standard algorithm for this application may not be optimal.

Our ODE solver is the standard RK4 stepper.  We experimented with a
6th order symplectic Runge-Kutta (using the \citealt{Yoshida:1990}
trick) and the Runge Kutta Felhberg algorithm
\citep[e.g.][]{Cash:1985}, and the adaptive Bulirsh-Stoer (B-S) method
with a specified relative error \citep[e.g.][]{Press.Teukolsky.ea:92}.
The non-sympletic B-S scheme is much more efficient owing to the
multiple length scales of the halo/spheroid and disk.  Relative energy
changes of one part in $10^5$ and relative $L_z$ changes of one part
in $10^8$ are obtained with time maximum time steps of 0.0001.
However, an externally chosen fixed time-step using the standard RK4
yield the most rapid convergence of the Lyapunov exponents on average.

The RK4 time-step, $\delta T$, is chosen as follows.  We begin with a
time step selected using the typical n-body simulation algorithms: a
fraction of the characteristic dynamical time determined by either the
logarithmic derivative of the position or velocity.  Then, we solve
the equations of motion numerically to find the smallest ratio of
velocity to force.  We use 1/100 of this value to integrate the
equations of motion and its tangent space to determine Lyapunov
exponents.  This fraction is both increased and decreased to check for
convergence of the derived exponents.  Typically, this procedure
results in a time step one order of magnitude smaller than that chosen
by the traditional algorithm used for n-body simulations.  In addition
to the ODE-solver time-step, one must choose the number of individual
time-steps between Gram-Schmidt orthogonalisation of the principle
directions.  After some experimentation, we choose this number to be
1/3 of the mean orbital time.  Our implementation was tested
successfully using the Lorenz equations which has well-known Lyapunov
values \citep{Wolf.etal:1985}.

Convergence of the maximal Lyapunov exponent, $\lambda$, for small
value of $|\lambda|$ can be logarithmically slow.  That is, if the
true value is $\lambda=0$, the most positive exponent approaches zero
proportional to $1/N$ where $N$ is the number of steps $\delta T$.
For the work here, we are less interested in the value of $\lambda$
but whether $\lambda>1/T$ for some characteristic time $T$.

Our algorithm is as follows:
\begin{enumerate}
\item Discard the first $N_{initial}$ of the recorded Lyapunov values.
  Typically, we choose a series $j\in N_{total}=2\times10^5$ steps of
  size $\delta T$ discard the first half of the series; that is, we
  set $N_{initial}=N_{total}/2$.
\item Order the exponents
  $\lambda_{0,j}>\lambda_{1,j}>\cdots>\lambda_{5,j}$.  For
  $\lambda_0$, compute the linear regression of $\log\lambda_{0\,i}$
  with $\log j$ to estimate the slope $m$.
\item Perfect convergence to zero yields $|m|=\mbox{constant}={\cal
    O}(1)$ while perfect to a non-zero positive value yields $m=0$.
  If the computed slope $m$ is greater that $-1/2$ we assume that the
  most positive exponent is zero.
\end{enumerate}

The phase space vector is computed from the background potential for
each torus in the grid (as described in section \ref{sec:toricomp}). In
this paper, we are typically analysing of order $10^4$ different
initial conditions for multiple numerical experiments, so examination
of each by eye is not feasible.  However, we have done this for some
test cases and failure modes in this algorithm warrant some discussion
because they appear to indicate interesting dynamical situations.
First, we noticed that the run of individual $\lambda_k$ with step $j$
do not maintain their initial ordering.  For example, it is not
unusual for the value of $\lambda_k$ to converge a different rates.
Secondly, and more importantly, individual orbits that appear to be
converging to positive value $\lambda_0=\epsilon>0$ at small $j$ will
begin to converge to zero logarithmically at large $j$.  Occasionally,
the opposite behaviour is seen: a value of $\lambda_0\propto 1/j$ at
small $j$ may suddenly increase and converge to decreasing value of
$\lambda_0=\epsilon>0$ at large $j$.

After checking for and not finding any implementation errors, we
concluded that this behaviour is consistent with the dynamics under
study.  In particular, an orbit in a chaotic region will exhibit
exponential sensitivity.  However, the apparent diffusion in action
space may result in an orbit ``sticking'' in a region of apparent
regularity for some time, resulting in a decreasing positive value of
$\lambda_0$.  This well-known behaviour is the result of an initially
diffusing orbit encountering regular islands around separatricies and
at the boundary of stochastic sheaths.  Leaving a detailed study for
later, we note that this tendency towards stickiness in stochastic
layers surrounding separatricies of strong resonances undermines the
utility of using Lyapunov exponents to characterise the dynamics in
these stochastic layers.  Also note that these same orbits that
exhibit non-uniform convergence often coincide with values of broken
tori identified using the numerical KAM method, suggesting that these
are regions of weak chaos (i.e. sub-exponential sensitivity) that
occur near strong resonances. Phenomenologically, these weakly-chaotic
trajectories appear regular throughout much of their orbital phase but
may switch morphology suddenly near heteroclinic points.  So, although
they may appear to have Lyapunov exponents close to zero, they give
rise to orbits with varying behaviour.

\section{Interpretation of broken tori}

Hamilton's equations for a Fourier-series perturbation with a single
dominant term (i.e. a single $\varpi_\bl$ in equation \ref{eq:w1aa} or
\ref{eq:Wexp}) may be solved directly without difficulty.  However,
difficulties do arise when several terms contribute comparable power
to equation (\ref{eq:w1aa}).  For example, consider two such terms
($A$ and $B$) with $\bl_A\not=\bl_B$.  As described in section
\ref{sec:HJ}, each term alone is analogous to a single pendulum with a
single degree of freedom.  Consider an initial condition near the
unstable equilibrium of one of the pendula.  The initial conditions
must be finely tuned for the trajectory to `climb' to the unstable
equilibrium.  As the trajectory approaches this equilibrium, the force
of the primary pendulum vanishes, and then, the other pendulum can
influence the trajectory, causing apparent stochastic behaviour.
Depending on the state of the full system, this influence can cause
the trajectory to enter the either the positive or negative rotation
zones or the libration zone of the primary pendulum.  In summary, the
two interacting pendula have resulted in the destruction the torus
described by the action conjugate to the angular degree of freedom
describing the unstable equilibrium.  In some cases, for very strong
perturbations, resonances from harmonics of the primary resonance can
interact.  This dynamics leads to the \emph{resonance overlap
  criterion} \citep{Chirikov:79}.  This will be familiar in
solar-system context as orbit instability induced by the overlap of
mean-motion resonances.

The consequences of this broken torus depends on the problem at hand.
If one of the pendula corresponds to a primary resonance, noise near
the resonance can cause the ensemble of trajectories representing the
entire torus (e.g. initial conditions with the same initial actions
but uniformly spaced phase values) to change their actions depending
on their phase.  Moreover, the problem can be complicated by many
Fourier terms simultaneously interacting and generating wide regions
in phase space that exhibit exponentially sensitive stochastic
behaviour.  The erratic behaviour can be dramatically exacerbated if the
unstable trajectories generated by several Fourier terms coincide in
phase space.  In this case, trajectories may move from resonance to
resonance.  In some cases, the influence of individual components may
be too plentiful to discriminate.  As previously described (see
section \ref{sec:lyapcompute}), the degree of stochasticity is most
often characterised by the product of Lyapunov exponent, $\lambda$ and
the characteristic dynamical time, $\tau$.  Values of
$\cC\equiv\lambda\tau\simeq1$ are called \emph{strong} chaos and
values of $\cC\lta1$ are called \emph{weak} chaos
\citep[e.g.][]{Zaslavskii.etal:1991}.

The method highlighted in section \ref{sec:KAM} is based on a
perturbative solution including all the lower-order terms $\bl$ within
some lattice with cardinality $\prod_jl_{j,max}$.  We assume that
failure to find a solution leading to a new torus implies a
\emph{broken} torus: the absence of an analytic predictive description
of future time evolution of an ensemble of nearby phase-space points.
Moreover, although broken tori and Lyapunov exponents are related, an
exploration of phase space using each method reveals different
information about the underlying dynamical structure.

For example, loci of broken tori in a phase space perturbed away from
regularity may result from a specific strong resonance.  Orbits of
different morphology reside on either side of or in islands
surrounding this resonance.  With increasing perturbation strength,
these loci broaden as additional terms conspire to break tori.
However, although a broad region of broken tori implies regions of
positive Lyapunov exponents, this does not imply that $\cC\gta1$, and
conversely, large $\cC$ may occur in narrow regions around
separatricies, as in the standard map.  As we will see below and in
\citetalias{Weinberg:15b} and \citetalias{Weinberg:15c}, regions of
high $\cC$ often appear in regions of broken tori but not exclusively.
In some cases, the largest values of $\cC$ occur between the widening
loci of primary resonances. The loci of original primary resonances
sometimes remains a local minimum in $\cC$, presumably owing to
regular islands \emph{spawned} by these resonances, as suggested by
the morphology of Poincar\'e surface-of-section plots for the standard
Hamiltonian mapping problem.

\subsection{A simple dynamical model}
\label{sec:toymodel}

\begin{figure}
\centering
\includegraphics[width=0.8\textwidth]{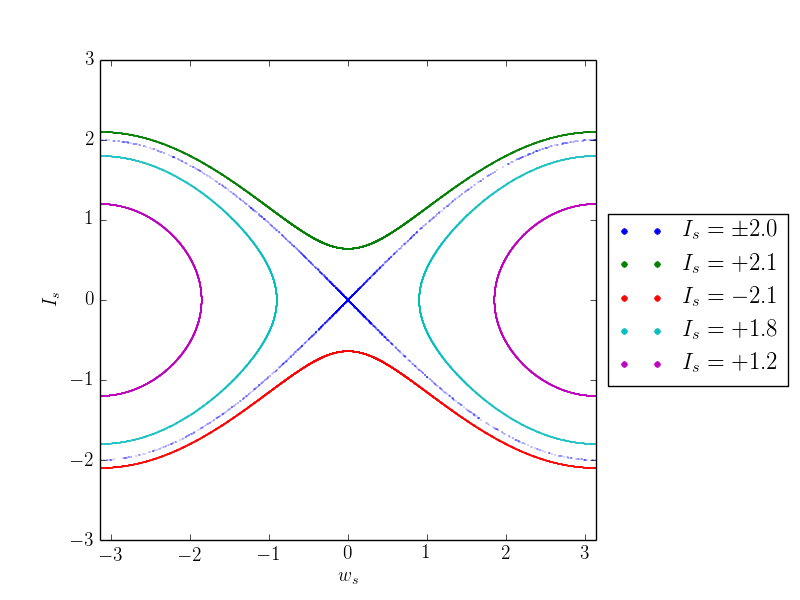}
\caption{Poincar\'e surface-of-section plot with $\varpi_A=1$ and
  $\varpi_B=0$ showing action $I_s$ versus angle $w_s$ for $w_f=0$ for
  approximately $10^4$ recurrences. The action of the homoclinic
  trajectory is $I_s=\pm2$.  Trajectories with $|I_s|<2$ and $|I_s|>2$
  are in libration and rotation, respectively. \label{fig:upert}}
\end{figure}

\begin{figure}
\centering
\includegraphics[width=0.8\textwidth]{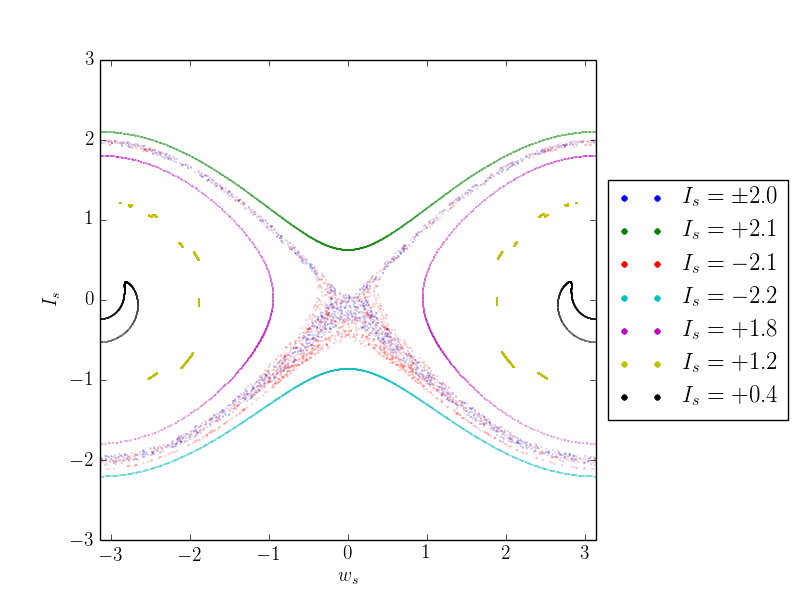}
\caption{As in Fig. \protect{\ref{fig:upert}} with
  $\varpi_B=0.01$. Although the amplitude of the second term is only
  1\% of the primary term, a region around the homoclinic trajectory
  is chaotic.  Also note the distorted resonant orbit for
  $I_s=0.4$. \label{fig:pert_0p01}}
\end{figure}

\begin{figure}
\centering
\includegraphics[width=0.8\textwidth]{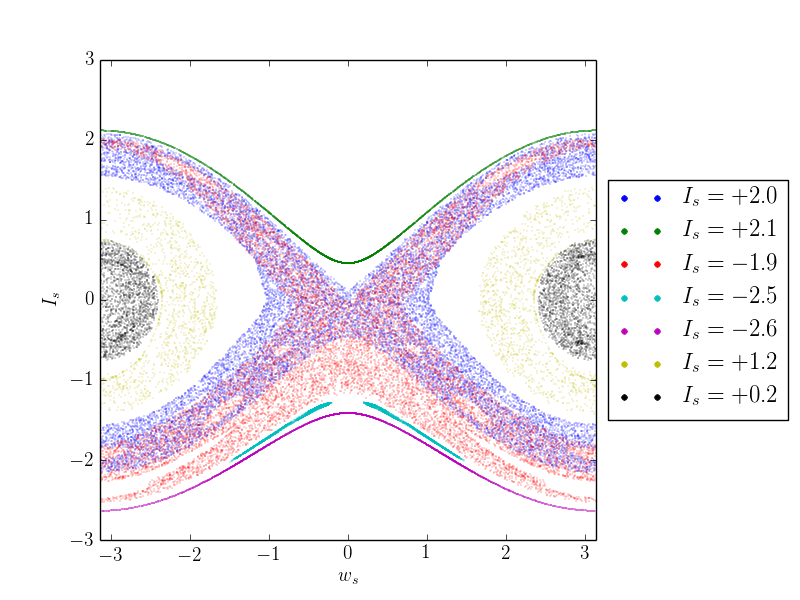}
\caption{As in Fig. \protect{\ref{fig:upert}} with $\varpi_B=0.1$. The
  amplitude of the second term is only 10\% of the primary term; most
  of the libration zone is filled with chaotic trajectories, although
  regular initial conditions can be found still.  For example,
  trajectories with $I_s=2.0, -1.9$ near the original homoclinic
  trajectory were chosen to display embedded islands.  An example of a
  resonant island trajectory is $I_s=-2.5$. \label{fig:pert_0p1}}
\end{figure}

In this section, we investigate the evolution of a two-pendulum model
with two non-zero terms in equation (\ref{eq:w1aa}) to explicitly
contrast the one-degree of freedom problem that underpins secular
evolution with the multiple degree of freedom problem that admits
chaos.  For an explicit example, let us consider the Hamiltonian
\begin{equation}
  H(\bI, \bw) = H_o(\bI) + \varpi_A(\bI) e^{i\bl_A\cdot\bw} +
  \varpi_B(\bI) e^{i\bl_B\cdot\bw},
  \label{eq:toyH}
\end{equation}
where $(\bI, \bw)$ are the action-angle variables for the unperturbed
Hamiltonian $H_o(\bI)$ with $\bI=(I_1, I_2)$ and $\bw=(w_1, w_2)$.
The restriction to a Fourier single term in the expansion series
results in classical secular perturbation theory.  Occasionally, the
linear solution described above may diverge owing to the resonant
denominators, but a non-linear solution can be found most often for a
single term.  However, when one or more \emph{additional} Fourier
terms are added to the Hamiltonian, irregularity arises.  The
importance of the irregularity to the solution in its astronomical
context depends on the relative magnitude of the expansion
coefficients and the order of commensurability.  We will quantify the
order of resonance as $||\bl|| \equiv \sup\{l_j\}$.

It is well-known that chaos arises in a \emph{stochastic layer} near
the homoclinic trajectory \citep[e.g.][]{Reichl:2004}.  So, we focus
on the solution of the perturbed Hamiltonian near the homoclinic
trajectory for a single perturbation term, say $\varpi_A\not=0,
\varpi_B=0$.  Motivated by orbit averaging, we transform to
\emph{slow} and \emph{fast} variables defined by the following Type 2
canonical transformation:
\begin{equation}
  F = (\bl_A\cdot\bw) I_s + w_1 I_f.
  \label{eq:toyF}
\end{equation}
Near the commensurability defined by $\bl_A\cdot\bw=0$, equation
(\ref{eq:Cw}) implies that the new angle $w_s=\bl_A\cdot\bw$ will
change slowly and the new $w_f=w_1$ will change rapidly with time.
The expressions for $I_s$ and $I_f$ follow similarly from equation
(\ref{eq:CI}).  The choice of the second term has a fair bit of
flexibility in this application; the most important constraint is that
$|{\dot w}_f|>0$ when $|{\dot w}_s|\approx0$.  These conditions can be
made rigorous \citep{Lochak.Meunier:1988} but are not needed for this
example.

To define an explicit form for the Hamiltonian in equation
(\ref{eq:toyH}), we expand $H_o(\bI)$ to second-order around a set of
unperturbed initial actions $(I_{s0}, I_{f0})$.  Ignoring the cross
term and completing the squares, we obtain:
\begin{equation}
  H(\bJ, \bw) = C(\bI_0) + 
  \alpha\frac{J_{s0}^2}{2} + \beta\frac{J_{f0}^2}{2} + 
  \varpi_A(\bI)e^{iw_s} + \varpi_B(\bI)e^{i(\gamma w_s + \delta w_f)}
  \label{eq:toyH2}
\end{equation}
where $C(\cdot)$ is an uninteresting constant and $\alpha$ and $\beta$
represent $\partial^2H/\partial J_s^2$ and $\partial^2H/\partial
J_f^2$ respectively, the definitions of $J_s, J_f$ include the offsets
from initial conditions and completing the squares, and $\gamma$ and
$\delta$ are $l_{B2}/l_{A2}$ and $l_{B1} - l_{B2}l_{A1}/l_{A2}$,
respectively.

For a simple example, we begin with initial conditions near the
homoclinic trajectory defined by $\alpha=1$, $\beta=1/2$, $\varpi_A=1,
\varpi_B=0$ with $\bl_A=(-1, 1)$ and $\bl_B=(1, 1)$.  This implies
that $\gamma=1$ and $\delta=2$.  The $\varpi_B=0$ surface-of-section
(SOS) plot (Fig. \ref{fig:upert}) reveals that standard phase space of
the regular, integrable pendulum.  For $\varpi_B>0$, the homoclinic
trajectory broadens owing the action of the non-resonant term on on
the resonant term near the unstable point.  The SOS plot for
$\varpi_B=0.01$ and $\varpi_B=0.1$ are shown in
Figs. \ref{fig:pert_0p01} and \ref{fig:pert_0p1}, respectively.
Typical values of $\varpi_B/\varpi_A$ for astronomical perturbations
described below and in the companion papers are between 0.1 and 0.5 so
Fig.  \ref{fig:pert_0p1} is the more typical situation.  For values
$\varpi_B/\varpi_A={\cal O}(0.1)$, much of the libration is chaotic
with a relatively wide chaotic zone around the original homoclinic
trajectory.  The dynamics of models like this are nicely described in
chapters 2 and 3 of \citet{Zaslavsky:2007}.

The model of equation (\ref{eq:toyH2}) is today example only; most
realistic expansion series have approximately 5 terms with amplitude
ratios with the primary between 0.1 and 1.0.  Many may interact
simultaneously leading to complex chaotic dynamics near the primary
resonant orbit.  This simple model helps motivate the prevalence of
chaotic behaviour in the astronomical examples in the companion papers.
Additional investigation of the dynamics of this simple model reveals
a wide variety of features found in the standard Hamiltonian mapping,
\begin{equation}
\theta_{n+1} -2\theta_n + \theta_{n-1} = -K\sin\theta_n
\label{eq:stdmap}
\end{equation}
with $K>1$.  In particular, one finds a stochastic `sea' with a wide
variety of islands whose location change quickly with small changes in
action within the stochastic layer.  This leads to the behaviour of
Hamiltonian intermittency: the appearance of both stochastic and
regular behaviour in the same trajectory.  This results from
trajectories being held up at the border of stable islands, sometimes
called ``stickiness'' \citep[i.e.][]{Lee:88,Lai:99}. These dynamics
may lead to heavy-tailed escape distributions similar to those caused
by the first-order Fermi effect \citep{Zaslavskii.etal:1991}.  The
relevance of possibly pervasive weak chaos in shaping the galactic
velocity distributions and its relevance to so-called radial migration
motivates additional investigation.

\section{Example: circular restricted three-body problem}

\label{sec:3body}

\subsection{Introduction}

The three-body problem is one of the oldest and best studied in
dynamics, although surprises continue to emerge.  Here, we will
consider application of the numerical KAM method to the restricted
three-body problem.  The main goal for this example is the comparison
our diagnosis of broken tori using the nKAM method with the
characterisation of chaos using Lyapunov exponents for a familiar
problem.

We take units $GM_\odot=1$, consider the perturbation to be a
circularly orbiting Jupiter-like planet with $M_p=0.001$ and $a=1$,
and explore the motion of a coplanar test particle.  We examined two
types of perturbations.  In the first, we assumed that the unperturbed
potential is that of a central point mass fixed at the origin
(i.e. the Sun) and the perturbation is an orbiting point mass
(i.e. Jupiter).  In the second, we assumed the perturbation is the
difference between the unperturbed potential and that generated by the
full two-body solution including barycentric motion.  The results were
nearly the same for both perturbations for harmonic orders $m=1$
through $m=3$, so we will use the fixed barycentre perturbation for
the remainder of the calculations.

\begin{figure}
  \includegraphics[width=0.8\textwidth]{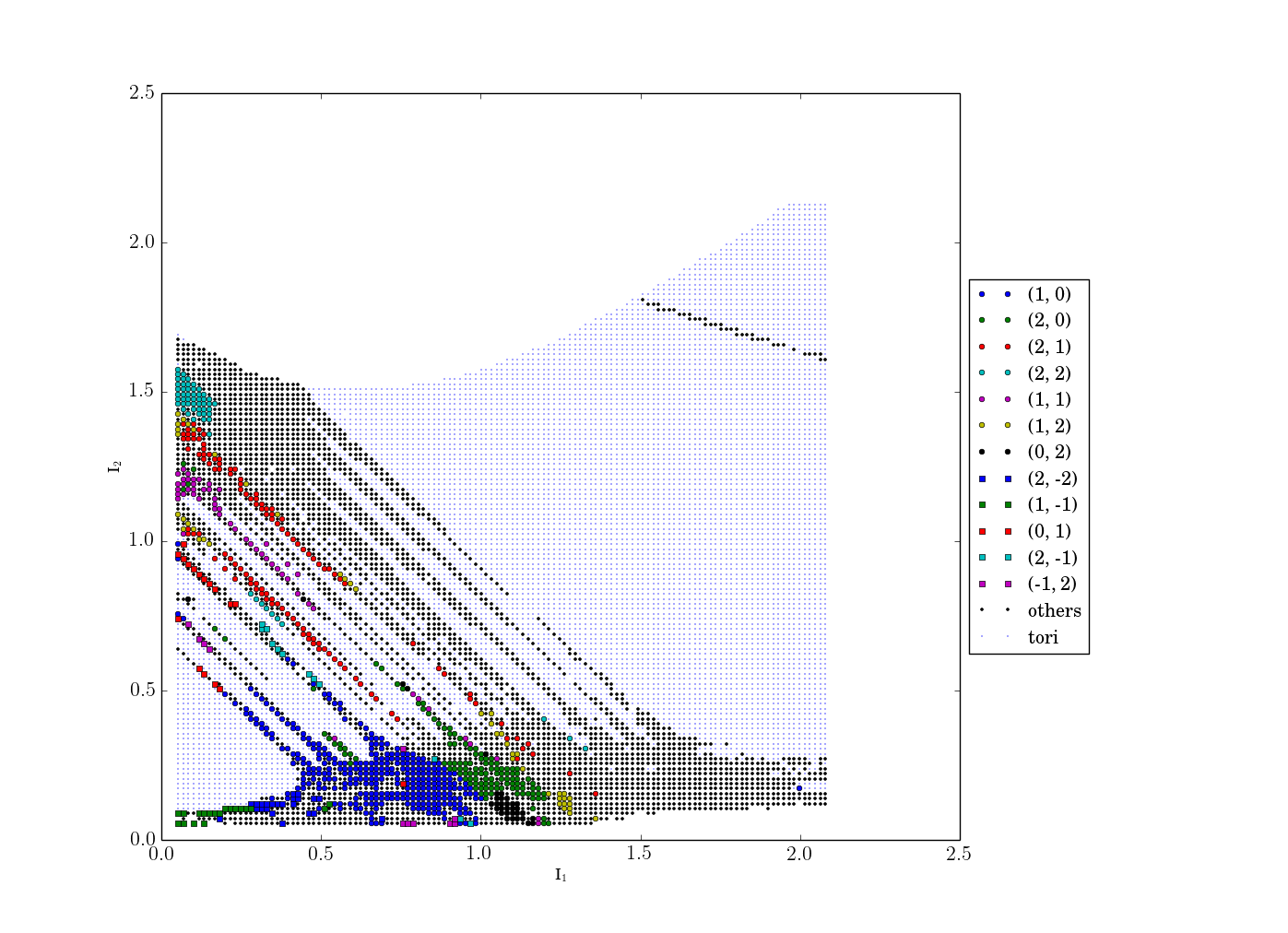}
  \caption{The broken and unbroken tori using the nKAM method for
    $|l_1|\le8, |l_2|\le8$ and azimuthal harmonic orders
    $l_3=m=0,\ldots,3$ in the action plane. The legend indicates
    primary components $W_{\bl}$ in the form $(l_1, l_2)$ for
    $|l_j|<3$, broken tori of higher order ($|l_j|\ge3$) are indicated
    by black dots and unbroken tori are indicated by blue
    dots.  \label{fig:jup_tori}}
\end{figure}

\begin{figure}
  \includegraphics[width=0.8\textwidth]{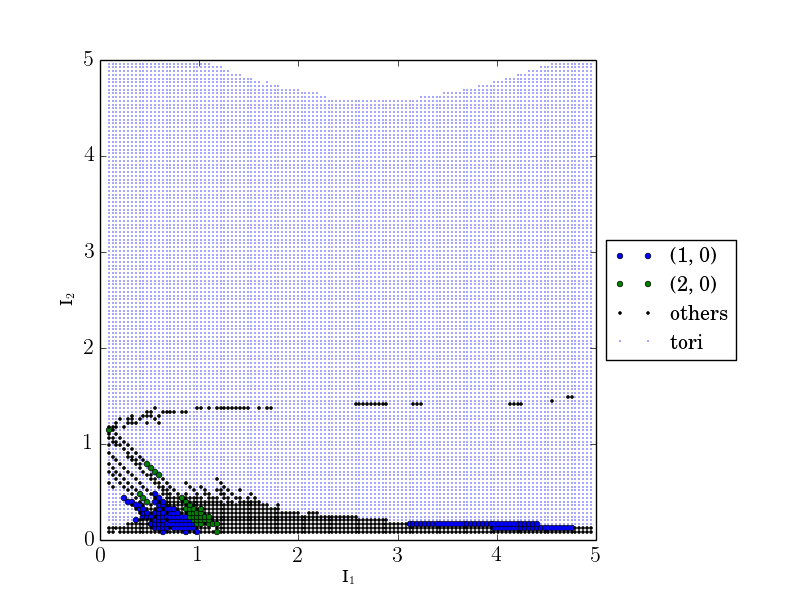}
  \caption{As in Fig. \protect{\ref{fig:jup_tori}} but for an action
    grid with larger maximum energy (larger semi-major axes), showing
    that broken tori are localised to the vicinity of the planet.
    \label{fig:jup_tori_big}}
\end{figure}

\begin{figure}
  \includegraphics[width=0.8\textwidth]{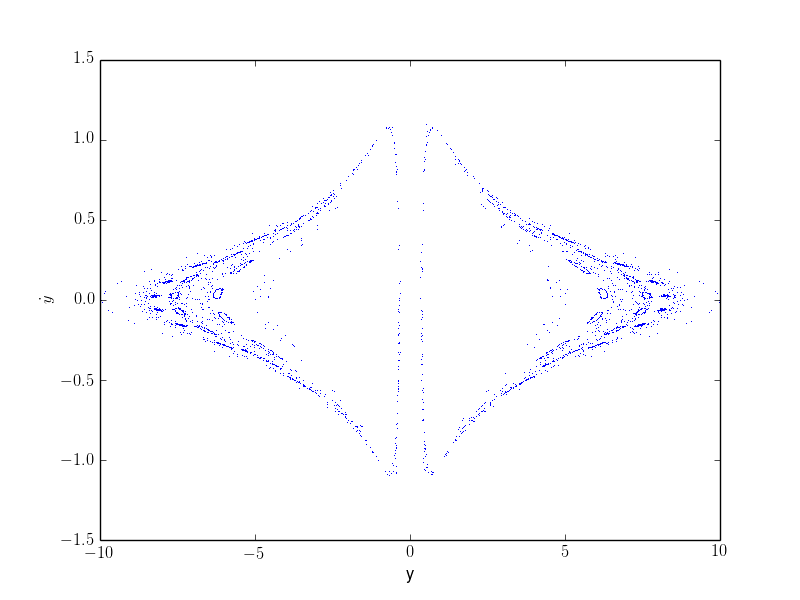}
  \caption{SOS plot for broken torus in upper-left most corner of the
    3:1 resonance locus in Fig. \ref{fig:jup_tori} with $(I_1, I_2) =
    (1.132, 0.837)$. \label{fig:sos31}}
\end{figure}

\subsection{Tori computation}

Here, we describe the results of the torus construction outlined in
section \ref{sec:KAM}.  Fig. \ref{fig:jup_tori} shows a plot of broken
(coloured circles and squares) and unbroken (blue dots) tori as a
function of radial and azimuthal actions, $I_1$ and $I_2$, for the
Jupiter-like model.  The original grid of radial and azimuthal actions
is chosen by choosing a range of minimum and maximum energy defined by
the circular orbits between $r_{min}=0.05$ and $r_{max}=120.0$.  Let
$E(r_c)$ and $J_{max}(E(r_c))$ be the energy and angular momentum of
the circular orbit at $r_c$.  For each energy value in the range
$[E(r_{min}), E(r_{max})]$, we compute the minimum and maximum values
of $I_1$ and $I_2$ such that $0.001<I_2/J_{max}(E)<0.999$.
Phase-space points outside the envelope of this region are shown as
white.  The legend has been restricted to low-order resonances for
clarity.  Other broken tori, for higher-order tori, are shown as black
points.  To help interpret the results, note that the two-body
(Kepler) problem in the adopted units has the following simple
representation as a function of $\bI$: the energy is
\begin{equation}
  H = E = -\frac{1}{2}\frac{1}{(I_1+I_2)^2},
  \label{eq:KepH}
\end{equation}
the eccentricity is
\begin{equation}
  \epsilon = \sqrt{1 - \bigg(\frac{I_2}{I_1+I_2}\bigg)^2},
  \label{eq:KepE}
\end{equation}
and the semi-major axis is:
\begin{equation}
  a = \bigg(I_1+I_2\bigg)^2.
  \label{eq:KepA}
\end{equation}
Therefore, diagonal lines at constant $I_1+I_2$ are lines of constant
energy and semi-major axis.  Loci of constant eccentricity are the
lines $I_1 = [(1-\epsilon^2)^{-1/2} - 1]I_2$; the diagonal $I_1=I_2$
has eccentricity $\epsilon\approx0.866$.  Because the frequencies are
independent of $\epsilon$ at fixed $E$, these diagonals are also lines
of constant orbital frequency and therefore loci of resonances as
well.  Initial phase-space values were selected from a grid of $I_1$
and $I_2$ values chosen between upper and lower energy bounds.  The
finite steps in each action value yields non-rectangular boundaries in
the $I_1$--$I_2$ plane.

The method described in section \ref{sec:canonical} is most efficient
when the Fourier expansion series converges quickly, that is, when the
value of $l_{max}$ is small.  However, for this planetary
perturbation, orbits with increasingly large $\epsilon$ can pass close
to the perturbing planet for increasingly large values of $I_1$ (or
energy $E$).  Convergence of the $W_1$ series would require increasing
large values of $l_{max}$.  Clearly for modest fixed values of
$l_{max}$, this method will only produce accurate tori estimates for
$I_2\gg I_1$.  Nonetheless, our example below will show that we can
still obtain interesting insight into the dynamics with a small finite
cube of $\bl$ pairs.

A number of features are immediately evident.  First, the broken tori
cluster around loci corresponding to principle low-order resonances.
We identify the coefficient $W_{\bl}$ with the largest magnitude as
the principle resonance of each torus; the low-order principle
resonance ``quantum'' numbers $(l_1, l_2)$ are indicated in the
legend.  The dominating terms in the series should reflect the
underlying drivers of the perturbation, however, the identification of
principle resonances is solely based on amplitude in
Figs. \ref{fig:jup_tori} and \ref{fig:jup_tori_big}.  In many cases,
the lowest-order term is only slightly smaller than the largest term
with higher-order $\bl$. Therefore, the identifications in this figure
may not precisely the morphological character of the broken torus but
still provide some clues to the important features of the dynamics.
Two or more terms $W_\bl$ of comparable amplitude are most likely
necessary for stochasticity.  If desired, one may further study the
details of specific broken tori using the direct integration approach
described in section \ref{sec:toymodel}.

The tori with $a<1$ unbroken except at high eccentricity,
$\epsilon\gtrsim0.95$.  For reference, the locus $I_2>4I_1$ have
$\epsilon<0.6$ and $I_2=I_1$ has $\epsilon=0.866$ (from equation
\ref{eq:KepE}).  The swath of broken tori with $I_1>0.5$, $I_2<0.3$
have been spot checked by direct integration and many of these do
indeed exhibit unusual stochastic behaviour ``by eye'' in the form of
eccentricity changes and sometimes irregular precession.  The primary
resonance identification suggests that resonance spreads out in to the
cloud, presumably aided by other high-amplitude subdominant terms
$W_{\bl}$.  The locus of broken tori in the upper right of the diagram
are induced by the 3:1 and 4:3 resonances and other mean-motion
harmonics with other high-order secondary resonances.  The SOS in
planet's rotating from for the upper-left most torus in this locus is
shown in Fig. \ref{fig:sos31}.

For comparison, Fig. \ref{fig:jup_tori_big} shows a grid of tori at
lower action-space resolution but larger maximum energy and radius.
This figure shows that the damage to the tori by the Jupiter-like
planet is localised to the vicinity of the planet and eccentric planet
crossing orbits, perhaps as expected.  The upper locus is, again, the
result of the 3:1 and 4:3 primary resonances.  Broken tori in this
locus are affected by significant stochastic behaviour as evidenced by
direct integration.

Direct integration broken tori at small $I_1$ and $0.5<I_2<1.7$ show
less dramatic stochastic behaviour.  There are a large number of
distinct groups of broken tori in Fig. \ref{fig:jup_tori} that might
be of interest for various dynamical situations.  Most of these are
likely to have been investigated previously by planetary dynamicists.
Our main goal, here, is to make contact with a well-established
problem.  Key to this, is understanding whether the failure of the
numerical KAM procedure to produce a torus implies an irregular,
stochastic orbit.  To better assess stochasticity, we turn to Lyapunov
exponent computation using the methods described in section
\ref{sec:lyapcompute}.

For direct confirmation of broken tori, the Melnikov method has become
the standard tool for detecting splitting of invariant manifolds for
systems of ordinary differential equations close to integrable ones.
The Melnikov method for two degree of freedom Hamiltonians described
by \citet{Holmes.Marsden:82}.  Their method estimates the magnitude of
the chaos by measuring the fraction of the energy surface where
chaotic motion occurs.  This idea will be explored in a future
contribution.

\subsection{Lyapunov comparison}

\begin{figure}
  \includegraphics[width=0.8\textwidth]{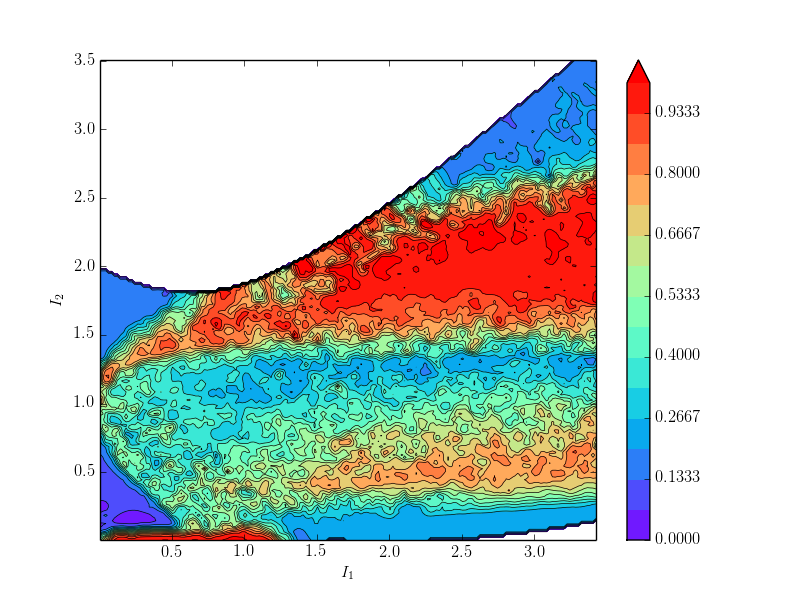}
  \caption{Density of action points from
    Fig. \protect{\ref{fig:jup_tori}} with positive Lyaponov exponents
    determined using the algorithm described in section
    \ref{sec:lyapcompute}. The data is somewhat under-smoothed to
    better reveal the fine-scale structure.\label{fig:lyap_o12}}
\end{figure}

\begin{figure}
  \includegraphics[width=0.8\textwidth]{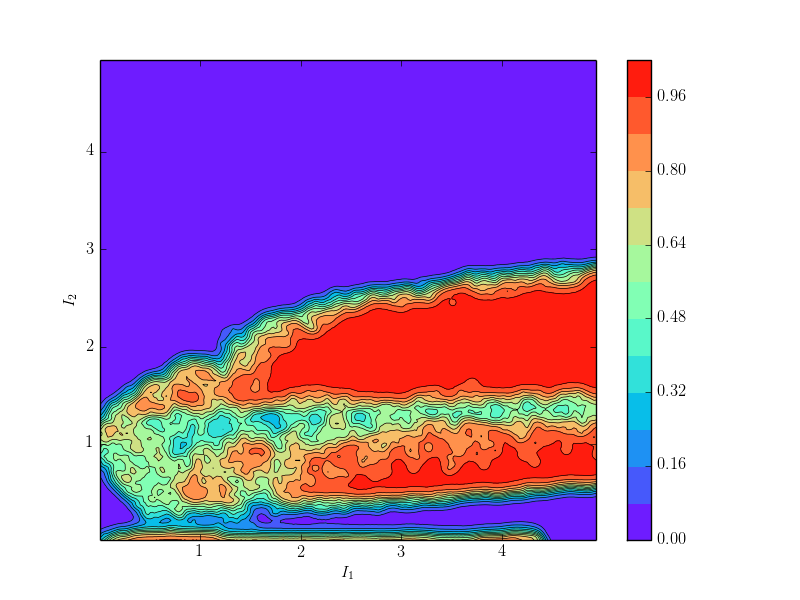}
  \caption{As in Fig. \protect{\ref{fig:lyap_o12}} using the action
    points from
    Fig. \protect{\ref{fig:jup_tori_big}}. \label{fig:lyap_o13} }
\end{figure}

\begin{figure}
  \includegraphics[width=0.8\textwidth]{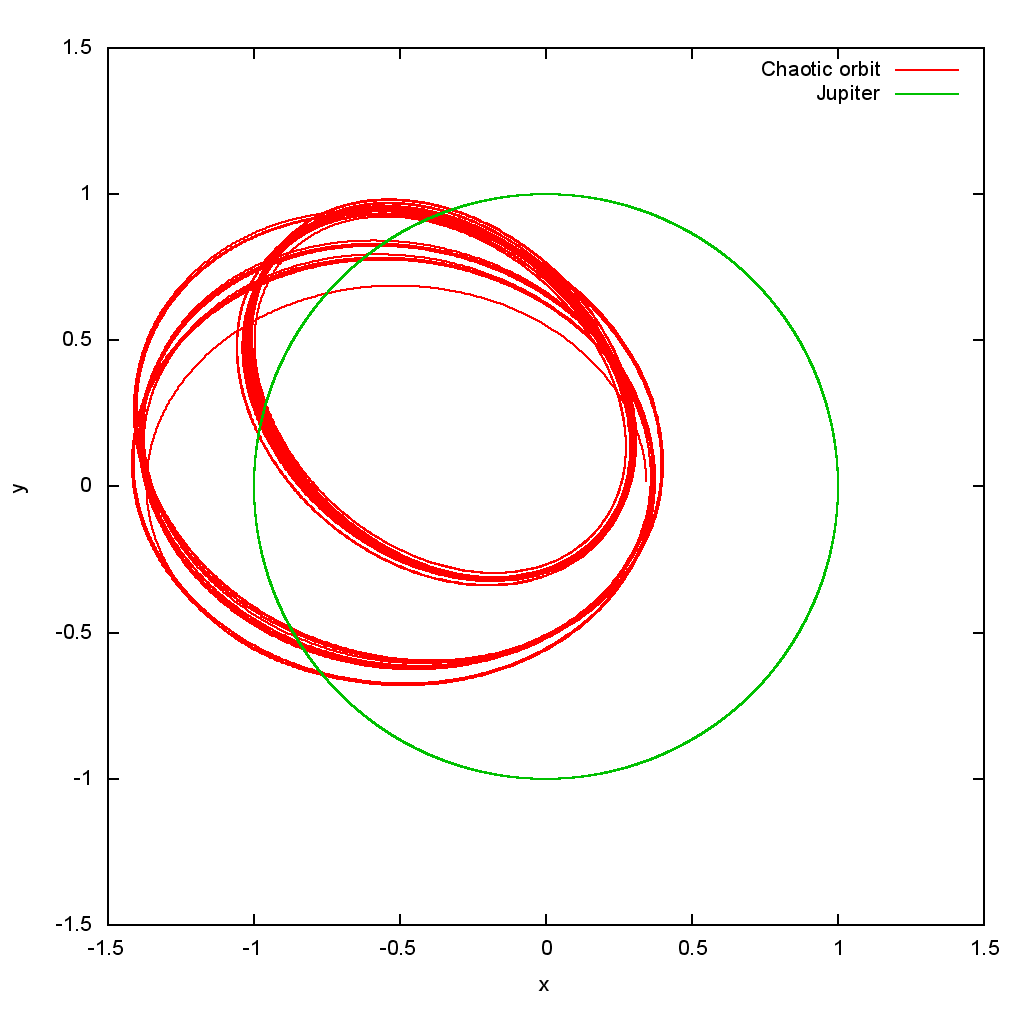}
  \caption{A chaotic orbit with $I_1= 0.185$ and $I_2=0.742$ (red).
    See Fig. \protect{\ref{fig:jup_tori}} to locate this broken torus.
    The orbit of the Jupiter-like planet is also shown (green).  The
    primary broken torus has commensurability $l_1=1, l_2=0, l_3=1$,
    with secondary terms $l_1=1, l_2=-1, l_3=2$, and $l_1=1, l_2=1,
    l_3=0$. \label{fig:orbit_chaotic} }
\end{figure}

We use initial conditions generated from the unperturbed tori in
Fig. \ref{fig:jup_tori} to estimate positive Lyapunov exponents using
the method outlined in section \ref{sec:lyapcompute}.  To recall, the
Lyapunov exponents for each orbit are computed using the tangent-space
method described by \citet{Christiansen.Rugh:97}.  To diagnose
convergence, the time series of exponent estimates, computed from a
running finite-time average, are analysed using linear regression
computation.  If the largest exponent approaches zero with a slope
$m>-1/2$, the exponent is assumed to be zero; otherwise, it assumed to
be positive.  The relative density of positive Lyapunov exponents
indicating stochasticity are shown in Fig. \ref{fig:lyap_o12}.  A
value of 1 (0) indicates that all (no) neighbouring orbits have
positive Lyapunov exponent. This figure is computed by smoothing the
irregular distribution of initial $(I_1, I_2)$ values onto a regular
grid with spacing 0.01 using a Gaussian kernel with width 0.01 in
action value.  The initial smoothing value was estimated by a
cross-validation analysis \citep{Silverman:86} and manually adjusted
downward by a factor of 5 to better reveal some of the intrinsic
structure common to Fig. \ref{fig:jup_tori}.

The comparison of Figs. \ref{fig:jup_tori} and \ref{fig:lyap_o12}
reveals both similarities and differences.  Overall, the regions of
broken tori coincide with regions of positive Lyapunov exponents.  The
3:1 resonance locus at $I_1\approx I_2\approx 1.8$ also appears as a
region of strong stochasticity.  Not present in
Fig. \ref{fig:jup_tori} are the horizontal loci of high-density
positive Lyapunov exponent with $I_1\approx1.4$ and $I_2>0.5$.
Additional investigation revealed these are caused by high-order
secondary resonances not included in the $W_\bl$ series.  These are
eccentric trajectories that come close to the planet at pericentre;
their inherent stochasticity would only be revealed by the nKAM method
for large $||\bl||$.  Similar a horizontal locus with $I_2<0.5$
\emph{is} present in Fig. \ref{fig:jup_tori}.  The direct integration
of the equations of motion for a particular initial condition in the
broad region of broken tori is shown in Fig. \ref{fig:orbit_chaotic};
changes in the argument of periapsis and eccentricity results from
close encounters with the perturbing planet, as expected.

Figs. \ref{fig:jup_tori_big} and \ref{fig:lyap_o13} compare the same
model with lower action resolution but larger maximum energy, large
radii orbits.  This further illustrates the major features of the
findings here.  First, the low-order resonances give rise to
stochasticity spreading from these resonances in the diagonal region
described by $0.5\lta I_1+I_2\lta 1.5$.  Second, these stochastic
zones spread for high eccentricity orbits ($I_1>I_2$) forming a
horizontal band of chaotic orbits with $I_2\approx0.3$. This region
has eccentricity $\epsilon\gtrsim0.9$.  This band results from
low-order terms interacting with high-order terms.
Fig. \ref{fig:lyap_o13} reveals an additional band at larger $I_2$,
`parented' by higher orders resonances.

The locus of broken tori parented by the 3:1, 4:3 and other harmonics
primary appears to demarcate the the upper horizontal tongue of
high-eccentricity ($I_1\gg I_2$) positive Lyapunov exponents.  As in
the case of Fig. \ref{fig:lyap_o12}, the lower horizontal loci have no
detected counterpart in the numerical KAM results, presumably owing to
the truncation in $||\bl||$ as previously discussed.  The region of
positive Lyapunov exponents is bounded from above by the locus of
eccentric orbits with pericentre values of $\approx 3$ and bound from
below by the locus with apocentre values of $\approx 0.5$.

Overall, we find that the nKAM method and the Lyapunov exponent
analysis provide a consistent picture of the broken tori in the
restricted three-body problem for a Jupiter-like planet.  The nKAM
method provides useful dynamical details, such as which resonances are
responsible for the stochastic behaviour around the separatrix.  This
is done by recording the dominant terms in the series defining the
perturbation (equation \ref{eq:w1aa}). These terms may be later used
later for detailed analyses, as desired.  The Lyapunov exponent
analysis gives a broad-brushed view of stochasticity and is not
restricted to low-order resonances.

\section{Discussion and summary}
\label{sec:summary}

\subsection{Main results}
\label{sec:mainsum}

This paper describes a method for estimating the existence and
destruction of regular orbits or \emph{tori} under a perturbation.
All orbits are assumed to be regular before applying a perturbation.
This is most easily realised by choosing a standard model in St\"ackel
form \citep{Stackel:1891,Eisenhart:1934,Binney.Tremaine:2008};
spherical and two-dimensional axisymmetric potentials are trivial
examples.  Classical Hamiltonian perturbation theory isolates the
response of individual resonances by identifying the particular
slowing varying degree of freedom identified with the commensurable
frequencies at resonance and averages over the rest.  However, this
procedure fails to include important dynamics near the resonance
(i.e. close to the separatrix of the principle resonance) owing to
non-negligible perturbations by the averaged terms that have been cast
away by the averaging.  The new method proposed here uses the
Hamilton-Jacobi equation to extend the reach of canonical perturbation
theory by including the mutual interactions of excited resonances.  It
is now well known that these terms lead to dynamical instability
through classical indeterminism or chaos
\citep[e.g.][]{Zaslavsky:2007}.  By retaining many terms in the
original series rather than dispatching them using the averaging
theorem, we develop a perturbation theory that includes chaos.

\subsection{Motivation and interpretation}

The new method is motivated by mathematical machinery used in some
proofs of KAM theory \citep[e.g.][section 3]{Poschel:2001} and their
correspondence to techniques used by this author previously.
Specifically, both the KAM procedure and the standard secular
perturbation theory attempts to solve the perturbed Hamiltonian
equation using a Fourier analysis.  The KAM procedure defines an
iterative procedure to obtain a new set of action-angle variables as
the solution to a perturbed Hamilton-Jacobi equation.  The standard
secular perturbation theory uses the averaging theorem to derive
one-dimensional equations of motion for each resonance of interest.
The essence of the method described in section \ref{sec:KAM} is a
quadratically converging iteration for the coefficients of a series
approximation to a canonical generating function that defines a new
set of action-angle variables under the perturbation for all terms
simultaneously.  That is, the converged coefficients give us a
canonical generating function that solves the perturbed
Hamilton-Jacobi equation and provides new action-angle variables
(i.e. new \emph{tori}).  Moreover, the KAM theorem suggests that it is
sufficient to retain some finite subset of coefficients for some
modest perturbation strength.  This justifies in part the truncation
of the expansion series that is necessary for practical numerical
computation.

When the coefficients for a particular value of the action vector,
$\bI$, fail to converge, the cause is rapid oscillation near the
unstable equilibrium of the primary resonance induced by one or more
secondary resonances. This is an example of the so-called homoclinic
tangle discovered by Poincar\'e. Specifically, the perturbation splits
the stable and unstable parts of the infinite-period resonant
trajectory; the frequency of their mutual intersections intersect
increase rapidly near the original unstable points of
equilibrium. \citet{Wiggins:1992} describes the dynamics of these
tangles and their consequences in exquisite detail.  From a heuristic
perspective, the KAM theorem tells us that the onset of chaos is a
result of coarse graining the dynamics generated by the homoclinic
tangles.  The consequences of these tangles is shown in the stochastic
layer that appears around the unperturbed homoclinic trajectory
(Fig. \ref{fig:upert}) as the amplitude of the perturbation is
increased (Figs. \ref{fig:pert_0p01} and \ref{fig:pert_0p1}) in our
toy model from section \ref{sec:toymodel}.  This observation also
suggests that the failure of the iteration in section \ref{sec:KAM} is
likely to imply chaos.  The identification of particular broken tori
provides a starting point for investigating the character and strength
of the resulting stochasticity.

The example in section \ref{sec:3body} shows that the numerical KAM
procedure reproduces many of the same features as the more traditional
empirical determination using Lyapunov exponents.  However,
characterising phase space using Lyapunov exponents is both time
consuming and numerically challenging.  They are also difficult to
connect to the underlying dynamics for several reasons.  First, the
computation of the Lyapunov exponents requires a numerical solution of
the equations of motion.  If the orbit is chaotic, the initial
conserved quantities drift.  Secondly, once the orbit enters the
vicinity of the homoclinic trajectory that has been broadened by
nearby resonances, the initially exponentially diverging orbits may
appear to converge again. In short, orbits with initially positive
Lyapunov exponents fall prey to the orbit's proclivity to explore
phase space and locate and get stuck near regular islands.  So
Lyapunov exponents are not ideal for classifying regularity in some
arbitrary potential owing to the appearance of islands of regularity
in the stochastic layers around primary resonances.  The same problem
affects Poincar\'e surface-of-section (SOS) plots; for a single orbit,
the SOS plot graphically represents the tendency for weakly chaotic
orbits to both diffuse to and from regular regions.  The numerical KAM
approach does not suffer from diffusion and is computationally much
faster that the standard Lyapunov exponent algorithms for studying the
phase space with the same number of initial conditions.  However, the
numerical KAM method requires a rapidly converging Fourier series for
accuracy, and this may not be satisfied for some problems of
astronomical interest.  For example, the example in section
\ref{sec:3body} illustrates that stochasticity for very eccentric
trajectories in a solar system requires expansion to very high order.
On the other hand, the applications described in \citet{Weinberg:15b}
and \citet{Weinberg:15c}---a bar perturbation to a disc and disc
perturbation to a dark-matter halo, respectively---will be well
described by a small number of terms.

In addition to analytic estimates of secular evolution and dynamical
stability, action variables are widely employed to represent
phase-space distributions by making use of Jeans' theorem
\citep{Binney.Tremaine:2008}.  Given a time-independent action-space
distribution, the positions and velocities of all orbits are known for
all time.  For many applications, the action-angle method will remain
a useful and mostly accurate construct.  The results of this paper,
combined with the galaxy applications described in
\citetalias{Weinberg:15b} and \citetalias{Weinberg:15c}, suggest that
systems with strong deviations from the classical regular systems
(spherical or triaxial St\"ackel potentials, two-dimensional
axisymmetric disks, etc.) may give rise to dynamics strongly affected
by, if not dominated by, irregularity.  For example, one would not use
Jeans' equations to describe the dynamics in the vicinity of Jupiter;
\citetalias{Weinberg:15b} and \citetalias{Weinberg:15b} describe
analogous situations for galactic dynamics.

A secondary goal of this paper is an extension of the well-established
tools of secular perturbation theory to the study of stochastic
dynamics in the context of galaxy evolution.  An important motivating
application not addressed here or in the companion papers is the
extension of this technique to the prediction of long-term evolution.
The might be done, for example, by using the numerical KAM procedure
to identify broken tori and then computing diffusion coefficients in
the initial action space following the numerical approach of
\citet{Chirikov:79} or something similar.  Phase-space evolution might
then be followed by a combination of the numerical KAM analyses and a
Fokker-Planck type evolution.  While feasible, this would be a
computationally expensive project.  

\subsection{Implications for n-body simulations}
\label{eq:nbody}

Some will no doubt argue that the applicability of the work described
here should be established in n-body simulations before proceeding
further.  This gives rise to the question of whether direct n-body
simulation suffices to obtain the dynamics described in this paper.
This is a separate project in itself but in aid of this goal, I offer
the following issues for consideration.

First, it is well established that stochastic layers exist around
primary resonances in Hamiltonian systems with multiple degrees of
freedom.  The width of these layers depend on the strength and
structure of the perturbation.  These layers are expected to have
lower dimensionality that the phase space itself.  Therefore, it is
possible that small time-steps, much smaller than typically used in
n-body simulations, might be necessary for individual particle orbits
to resolve (or ``feel'') the chaos.  In other words, large time-steps
may be giving the appearance of more regularity than in nature.
Anecdotal evidence for this issue is found in the computation of
Lyapunov exponents.  Large time-steps, say 20--100 divisions per
characteristic orbital time, are too large to capture the effect of
exponential growth owing to resonance overlap; time-steps 10--100
times smaller were required for convergence of the exponents.

Secondly, adaptive Poisson solvers generate force fluctuations.  Their
scale depends on the method.  For direct-summation codes and tree
codes, the spatial scale will be the smoothing length.  For PIC-type
solvers, the scale will be the cell size.  For orthogonal-basis codes,
the scale is set by the maximum order of the basis functions.  The
concern is that scattering from these fluctuations may interfere with,
change or otherwise wash out the underlying Hamiltonian chaos.  Some
will argue that galaxies are naturally noisy, and therefore, if the
numerical noise in the force computation washes out the Hamiltonian
chaos, then it is probably not important.  This argument is circular.
The existence of astronomical noise does not make n-body particle
noise a good stand in.  Even in the unlikely case that particle noise
emulated the true fluctuations for some application, the simulator
would need to compute the auto-correlation time for the noise
fluctuations to set the correct dynamical time-step.  

The natural noise in dark-matter halos will be a mixture of
particle-like and non-particle like.  Particle-like noise will come
from dark substructure and dwarf satellites, molecular clouds and star
clusters, to name a few.  However, it is likely that these are fewer
in number than simulation particles and with different distributions
than the dark matter or disk particles, and therefore the resulting
noise spectrum will be different.  Non-particle like noise will be
generated by the wrapping dark-matter streams, continuing gas
accretion and other long-term evolutionary processes that result from
galaxy assembly.  In short, all of these processes require precise
study before their contribution to the dynamics is well-understood.
 
\subsection{Topics for future work}
\label{sec:future}

Chaotic dynamics may give rise to anomalous acceleration or scattering
caused by overlapping patterns.  In the example from section
\ref{sec:toymodel}, the stochastic layer that forms around the
separatrix has multiple embedded islands.  Trajectories move through
the stochastic layer feeling the intermittent effects regular and
irregular motion.  For example, in computing the Lyapunov exponents
for section \ref{sec:3body} we noticed the tendency for trajectories
with positive exponents at early times begin to converge toward zero.
This is consistent with the behaviour of Hamiltonian intermittency.
This well-established behaviour is called \emph{stickiness}.  When
chaotic trajectories approach the regular regions they stick to their
border inducing long periods of almost regular motion. This
intermittent behaviour determines the magnitude of diffusion.  For
stronger perturbations, chaotic trajectories may appear to suffer from
noticeable anomalous kicks and apparently sudden changes in their
direction or orientation; such effects are related to ``radial
migration''.

In cases where the Lyapunov exponents are positive but very small, the
trajectories slowly diffuse into the thin chaotic layer defined by the
dominant and secondary term(s) in the action-angle expansion of the
perturbation.  Additional terms may yield a complicated network of
higher-order resonances, often sticking for very long times to the
boundaries of islands constituting the so-called \emph{edge of chaos}
regime \citep{Tsallis:2009}.  This seems less likely to be important
to time-limited systems such as galaxies.  A recent paper by
\citet{Manos.Machado:2014} notes that the measured fraction of regular
orbits in a barred galaxy simulation appears to increase with time.
Perhaps this is a consequence of intermittency.

\section*{Acknowledgments}

This early development of this work was supported in part by NSF
awards AST-0907951 and AST-1009652.  MDW gratefully acknowledges
support from the Institute for Advanced Study, where work on this
project began.  Many thanks Douglas Heggie who made valuable comments
on an early version of this manuscript.

\bibliographystyle{mn2e}

\appendix

\section{A computationally more efficient algorithm}
\label{sec:numerical}

The development of the nKAM algorithm in section \ref{sec:KAM} uses a
Fourier series in the new action-angle variables of the perturbed
Hamiltonian $(\bJ, \bs)$.  While elegant, this set of variables
complicated the solution for the new variables in terms of the old
variables $(\bI, \bw)$.  Rather than solve equation (\ref{eq:HJq}) by
a Fourier series in $\bs$ as described by equation (\ref{eq:ql1}), we
may write the expansion in terms of the old angles $\bw$:
\begin{equation}
  W_1(\bJ,\bw) = \sum_{\bl}
  \varpi_{\bl}(\bJ) e^{i\bl\cdot\bw}
  \label{eq:Wexp}
\end{equation}
where ${\bf J}$ is the new action vector, ${\bf w}$ is the old angle
vector, and $\bl=(l_1, l_2, l_3)$ is a vector of integers whose values
are $|l_j|\in{0, l_{max}}$ for some $l_{max}>0$ as before.  This mixed
set of variables is more natural for the Type 2 generating function
\citep{Goldstein.etal:02} but our main motivation here is
computational efficiency.  This mixed expansion series results in
iterative algorithm that does not require implicit solution of angle
variables.  The revised derivations otherwise follows the same steps
that lead to equation (\ref{eq:qquad}).

To see this new formulation facilitates the implicit solution of the
new canonical variables as a function of the old, write the expansion
of the right-hand side of equation (\ref{eq:HJ}) to first order in the
perturbation $W_1=\mathcal{O}[H_1]$ explicitly:
\begin{equation}
  H_{new}(\bJ) = H\bigg(\bJ+\frac{\partial W_1^{[n+1]}}{\partial\bw}, \bw\bigg)
  = 
  \bigg\{H_0(\bJ) +
    \bO(\bJ)\cdot\frac{\partial W_1}{\partial\bw} +
    H_1(\bJ, \bw)\bigg\} + \mathcal{O}(W_1^2)
  \label{eq:RHS}
\end{equation}
where we have used the notation $\Omega_j(\bJ) = \partial H_0/\partial
I_j|_{\bJ}$ for compactness.  If we subtract the terms in $\{\}$ from
both sides of equation (\ref{eq:RHS}), after rearranging and grouping
the new terms, we have:
\begin{eqnarray}
  H_{new}(\bJ) - H_0(\bJ) - \bO(\bJ)\cdot\frac{\partial W_1}{\partial\bw} - H_1(\bJ, \bw)
  &=&  
  \bigg[H_0\bigg(\bJ + \frac{\partial
        W_1}{\partial\bw}\bigg) - H_0(\bJ) - \bO(\bJ)\cdot\frac{\partial W_1}{\partial\bw}\bigg]
  \nonumber + \\ && 
  \bigg[H_1\bigg(\bJ + \frac{\partial W_1}{\partial\bw}, \bw\bigg) - 
    H_1(\bJ, \bw)\bigg].
    \label{eq:HJ2}
\end{eqnarray}
The left-hand side is the difference between $H_{new}(J)$ and the
first-order Taylor series.  Terms on the right-hand side of equation
(\ref{eq:HJ2}) are second order in the perturbation $V_1$.  As in the
section \ref{sec:derive}, we can exploit this to determine the
first-order contribution once and for all, and iterate to improve the
second-order contribution.

We proceed as follows. Consider the action-angle expansion of equation
(\ref{eq:HJ2}) in the Fourier series defined by equation
(\ref{eq:Wexp}).  The first two terms on the left-hand side are
constants with respect to $\bw$ and therefore only contribute to
coefficients $\varpi_{\bl}$ with $\bl=0$.  From equation
(\ref{eq:CI}), these terms make no contribution to the new action, and
from equation (\ref{eq:Cw}), these terms only result in a constant
offset in new angle.  As in section \ref{sec:KAM}, we see from
equations (\ref{eq:CI}) and (\ref{eq:Cw}) that the $\bl=0$ transform
of equation (\ref{eq:HJ2}) generates an uninteresting phase shift in
the solution.  The Fourier coefficient of the third term on the
left-hand side of equation (\ref{eq:HJ2}) is
$-i\bl\cdot\bO(\bJ)\,\varpi_{\bl}(\bJ)$ while that for the fourth term
is determined once and for all by direct quadrature.  Since the
right-hand side is second order in $W_1$, we can immediately solve for
the first-order contribution to $\varpi_{\bl}$ by demanding that the
first-order contribution to the LHS of equation (\ref{eq:HJ2})
vanishes:
\begin{equation}
  \varpi_{\bl}^{[1]}(\bJ)
  = \frac{i}{\bl\cdot\bO(\bJ)}\frac{1}{(2\pi)^3}
  \oint d\bw H_1(\bI, \bw) e^{-i\bl\cdot\bw}
  \label{eq:HJ1st}
\end{equation}
where $\bI=\bJ$ and $\bs=\bw$ to first order.  Again, we can safely
ignore the $\bl=0$ which yields an constant offset in $\bs$ by
equation (\ref{eq:Cw}).  We may now substitute this first-order
solution
\begin{equation}
  W_1^{[1]}(\bJ,\bw) = \sum_{\bl}
  \varpi_{\bl}(\bJ)^{[1]} e^{i\bl\cdot\bw}
\end{equation}
into equation (\ref{eq:HJ2}) to get:
\begin{eqnarray}
  \varpi_{\bl}^{[2]}(\bJ) &=& \varpi_{\bl}^{[1]}(\bJ) 
  +\frac{i}{\bl\cdot\bO(\bJ)}\frac{1}{(2\pi)^3}
  \bigg\{
    \oint d\bw\, \bigg[H_0\bigg(\bJ + \frac{\partial
          W_1^{[1]}}{\partial\bw}\bigg) - H_0(\bJ) -
      \bO\cdot\frac{\partial W_1^{[1]}}{\partial\bw}\bigg]
    e^{-i\bl\cdot\bw} + \bigg .
  \nonumber \\
  && \hspace{98pt}\bigg. \oint d\bw\,
    \bigg[H_1\bigg(\bJ + \frac{\partial W_1^{[1]}}{\partial\bw},
        \bw\bigg) - H_1(\bJ, \bw)\bigg]e^{-i\bl\cdot\bw}
  \bigg\}
\label{eq:HJ3}
\end{eqnarray}
Equation (\ref{eq:HJ3}) may be similarly iterated to obtain
\begin{eqnarray}
  \varpi_{\bl}^{[n]}(\bJ) &=& \varpi_{\bl}^{[1]}(\bJ) + \nonumber \\
  && \frac{i}{\bl\cdot\bO(\bJ)}\frac{1}{(2\pi)^3}
  \bigg\{
    \oint d\bw\, \bigg[H_0\bigg(\bJ + \frac{\partial
          W_1^{[n-1]}}{\partial\bw}\bigg) - H_0(\bJ) -
      \bO\cdot\frac{\partial W_1^{[n-1]}}{\partial\bw}\bigg]
    e^{-i\bl\cdot\bw} + \bigg .
  \nonumber \\
  && \hspace{98pt}\bigg. \oint d\bw\,
    \bigg[H_1\bigg(\bJ + \frac{\partial W_1^{[n-1]}}{\partial\bw},
        \bw\bigg) - H_1(\bJ, \bw)\bigg]e^{-i\bl\cdot\bw}
  \bigg\}
\label{eq:HJN}
\end{eqnarray}
or using $H=H_0+H_1$,
\begin{eqnarray}
 \varpi_{\bl}^{[n]}(\bJ) &=& \varpi_{\bl}^{[1]}(\bJ) + \nonumber \\
 && \frac{i}{\bl\cdot\bO(\bJ)}\frac{1}{(2\pi)^3}
  \bigg\{
    \oint d\bw\, \bigg[H\bigg(\bJ + \frac{\partial
          W_1^{[n-1]}}{\partial\bw}\bigg) - H(\bJ,\bw) -
      \bO\cdot\frac{\partial W_1^{[n-1]}}{\partial\bw}\bigg]
    e^{-i\bl\cdot\bw} \bigg\} \nonumber \\
\label{eq:HJN2}
\end{eqnarray}
The iteration may be terminated when the value of
$\max_{\bl}|\varpi_{\bl}^{[n]} - \varpi_{\bl}^{[n-1]}|$ is
sufficiently small.

Equations (\ref{eq:HJ1st}), (\ref{eq:HJ3}), (\ref{eq:HJN}), and
(\ref{eq:HJN2}) require angle integrals over a range of $\bl$.  The
restricted three-body problem example in section \ref{sec:3body} and
the galactic-bar perturbation in \citetalias{Weinberg:15b} considers a
two-dimensional orbital distribution with a non-axisymmetric
perturbation.  For the application in \citetalias{Weinberg:15c}, we
will consider azimuthally symmetric perturbations only; $L_z$ will
remain constant for all orbits and only terms $\varpi_{\bl}$ with
$l_3=0$ will be non-zero.  Therefore, our angle integrals in the
equations above are two dimensional for this case as well.  By
restricting our series to $|l_1|\in[0,l_{1,max}]$ and
$|l_2|\in[0,l_{2,max}]$, the angle integrals in equations
(\ref{eq:HJ1st}), (\ref{eq:HJ3}) and (\ref{eq:HJN}) have the same form
as a discrete Fourier transform and may be efficiently performed using
the FFT algorithm.  The two-dimensional restriction is for numerical
convenience only; see \citetalias{Weinberg:15c} for a discussion of
the full three-dimensional perturbation.

\section{Grid choice and interpolation}
\label{sec:grid}

As described in section \ref{sec:interp}, we divided our action grids
into three classes: 1) two-dimensional problems with action space
$(I_1, I_2$); 2) three-dimensional problems with action space $(I_1,
I_2, I_3)$ with one action conserved, $I_3=L_z$ say; and 3) the fully
general three-dimensional problem.  Our software implementation
includes all three classes but only (1) and (2) are used in this and
the companion papers.  In general, the distribution of $\bI$ will not
be on a regular grid, and therefore interpolation and evaluation of
partial derivatives require specialised methods. We have explored
spline fits, interpolating simplices \citep{Riachy.etal:2011} and the
Shepard method \citep{Shepard:68}.  Numerical experiments suggests
that the Shepard method is the most robust and generally accurate on
test problems.  The authors of ALGLIB (\url{www.alglib.net}) suggest
that the radial basis function (RBF) methods
\citep[e.g.][]{Buhmann:2003} further improve the performance of
Shepard's algorithm, but this has not been investigated here.

Our implementation of Shepard's algorithm begins with an irregular
point set denoted as $\bI_i$ for $i=1,\ldots,n$.  Let the desired
field $F(\bI)$ evaluated on this set be denoted as $F_i$.  The
original Shepard approximating function \citep{Shepard:68} is defined
as
\begin{equation}
  {\tilde F}(\bI) = \sum_{i=1}^n w_i(\bI) F_i
  \label{eq:shep1}
\end{equation}
where the weight functions $w_i$ are
\begin{equation}
  w_i(\bI) = \frac{H[\delta - d_i(\bI)]^r}{\sum_{j=1}^n H[\delta - d_j(\bI)]^r}
  \label{eq:shep2}
\end{equation}
with $\delta>0$, $d_i=||\bI - \bI_i||$, the standard Euclidean norm,
and $H[\cdot]$ defined by
\begin{equation}
  H[q] = 
  \begin{cases}
    $q$, & q\ge0\\
    $0$, & q<0\\
  \end{cases}
\end{equation}
The smoothing parameter $\delta$ and the exponent $r$ are tuning
constants.  Generally, if exponent $r$ has a value in $[0,1]$ then
${\tilde F}$ has peaks at the grid points, and if $r>0$, the
approximating function tends toward a well defined slope at the nodes.
In this paper, we use the modified Shepard method implementations by
\citet{Renka:88a,Renka:88b}.  This updated algorithm generalises the
simple weights of the original algorithm by augmenting the
approximation of $F_i$ by a quadratic multinomial whose coefficients
are determined by the weighted least-squares fit to the $N$ nearest
neighbours to $\bI$.  The weights are
\begin{equation}
  w_i(\bI) = \frac{H[r_i(N_q) - d_i(\bI)]^2}{[r_i(N_q) d_i(\bI)]^2}
  \label{eq:shep3}
\end{equation}
where $r_i(N_q)$ is the radius about $\bI_i$ containing $N_q$ points.
We used the recommended default value of $N_q=13$ and $17$ for the
two- and three-dimensional cases from \citet{Renka:88a,Renka:88b}.
The derivatives (e.g. for gradients of field quantities) are easily
computed analytically from the quadratically augmented form of $F_i$.
Finally, at each node in the grid, we store the full orbit
computation, tabulated as function of angle.  These are linearly as
needed to produce a full phase-space evaluation at $(\bI, \bw)$.

\label{lastpage}

\end{document}